\documentclass[preprint,tightenlines,aps,nofootinbib,showpacs]{revtex4}

\usepackage{dcolumn}
\usepackage{bm}
\usepackage[utf8]{inputenc}
\usepackage{graphicx}
\usepackage[colorlinks]{hyperref}

\usepackage{slashed}
\usepackage{epsfig}
\usepackage{color}
\PassOptionsToPackage{caption=false}{subfig}
\usepackage{subfig}

\definecolor{LightCyan}{rgb}{0.88,1,1}
\definecolor{piggypink}{rgb}{0.99, 0.87, 0.9}
\definecolor{applegreen}{rgb}{0.55, 0.71, 0.0}
\definecolor{darkpastelgreen}{rgb}{0.01, 0.75, 0.24}
\definecolor{green-yellow}{rgb}{0.68, 1.0, 0.18}
\definecolor{nicered}{rgb}{0.7,0.1,0.1}
\definecolor{nicegreen}{rgb}{0.1,0.5,0.1}
\definecolor{darkblue}{rgb}{0.0,0.0,.4}
\definecolor{darkred}{rgb}{0.4,0.0,0.0}
\definecolor{red}{rgb}{1.0, 0, 0}

\def\lsim{\mathrel{\rlap{\lower4pt\hbox{\hskip1pt$\sim$}}
     \raise1pt\hbox{$<$}}}         
\def\gsim{\mathrel{\rlap{\lower4pt\hbox{\hskip1pt$\sim$}}
     \raise1pt\hbox{$>$}}}

\def\beqn{\begin{eqnarray}} 
\def\eeqn{\end{eqnarray}} 
\def\be{\begin{equation}}
\def\ee{\end{equation}}

\def\alt{~\mbox{\raisebox{-.6ex}{$\stackrel{<}{\sim}$}}~}
\def\agt{~\mbox{\raisebox{-.6ex}{$\stackrel{>}{\sim}$}}~}

\newcommand{\tchi}{\tilde{\chi}}

\newcommand{\notE}{ \hbox{{$E$}\kern-.60em\hbox{/}}}
\newcommand{\notp}{\ \hbox{{$p$}\kern-.43em\hbox{/}}}
\def\eslt{\slashed{E}_T}
\def\to{\rightarrow}

\def\bi{\begin{itemize}}
\def\ei{\end{itemize}}

\def\tchi{\tilde\chi}

\def\tz{\tilde{\chi}^0}

\def\tg{\tilde g}

\preprint{
\hbox to \hsize{
\hskip0.1in 
\hskip3.6in $\vcenter{
                      \hbox{\bf OU-HEP-211202}
                      \hbox{December 2021}}$ }
}

\begin{document}


\title{
Detecting Heavy Higgs Bosons from Natural SUSY\\
at a 100 TeV Hadron Collider}


\author{Howard Baer}
\email{baer@nhn.ou.edu}
\affiliation{Homer L. Dodge Department of Physics and Astronomy, 
University of Oklahoma, Norman, OK 73019, USA}

\author{Vernon Barger}
\email{barger@pheno.wisc.edu}
\affiliation{Department of Physics, 
University of Wisconsin, Madison, WI 53706, USA}

\author{Rishabh Jain}
\email{rishabh.jain@hep1.phys.ntu.edu.tw}
\affiliation{Department of Physics, 
National Taiwan University, Taipei, Taiwan 10617, R.O.C}

\author{Chung Kao}
\email{Chung.Kao@ou.edu}
\affiliation{Homer L. Dodge Department of Physics and Astronomy, 
University of Oklahoma, Norman, OK 73019, USA}

\author{Dibyashree Sengupta}
\email{dsengupta@phys.ntu.edu.tw}
\affiliation{Department of Physics, 
National Taiwan University, Taipei, Taiwan 10617, R.O.C}

\author{Xerxes Tata}
\email{tata@phys.hawaii.edu}
\affiliation{Department of Physics, 
University of Hawaii, Honolulu, HI 96822, USA}

\bigskip

\date{\today}


\begin{abstract}

Supersymmetric models with radiatively-driven naturalness (RNS) enjoy
low electroweak fine-tuning whilst respecting LHC search limits on
gluinos and top squarks and allowing for $m_h\simeq 125$ GeV.  While
the heavier Higgs bosons $H,\ A$ may have TeV-scale masses, the
SUSY conserving $\mu$ parameter must lie in the few hundred GeV range.
Thus, in natural SUSY models there should occur large heavy Higgs boson
branching fractions to electroweakinos, with Higgs boson decays to
higgsino plus gaugino dominating when they are kinematically
accessible.  These SUSY decays can open up new avenues for discovery.
We investigate the prospects of discovering heavy neutral Higgs bosons
$H$ and $A$ decaying into light plus heavy chargino pairs which
can yield a four isolated lepton plus missing transverse energy
signature at the LHC and at a future 100~TeV $pp$ collider.
We find that
discovery of heavy Higgs decay to electroweakinos via its $4\ell$
decay mode is very difficult at HL-LHC.  For FCC-hh or SPPC, we study
the $H,\ A \to $ SUSY  reaction along with dominant physics
backgrounds from the Standard Model and devise suitable selection
requirements to extract a clean signal for FCC-hh or SPPC with
$\sqrt{s}=100$ TeV, assuming an integrated luminosity of 15 $ab^{-1}$.
We find that while a conventional cut-and-count analysis yields a
signal statistical significance greater than $5\sigma$ for
$m_{A,H}\sim 1.1-1.65$ TeV,
a boosted-decision-tree analysis allows for heavy Higgs signal discovery
at FCC-hh or SPPC for $m_{A,H}\sim 1-2$ TeV.
\end{abstract}

\maketitle

\section{Introduction} 
\label{sec:intro}

With the discovery of the 125 GeV Standard Model-like Higgs boson at
LHC~\cite{higgs}, all the particle states required by the Standard Model
(SM) have been confirmed.  And yet, many mysteries of nature still remain
unsolved.  Supersymmetric extensions of the SM are highly motivated in
that they offer a solution to the gauge hierarchy problem
(GHP)~\cite{ghp} which arises from the quadratic sensitivity of the
Higgs boson mass to high scale physics. SUSY models are also supported
indirectly by various precision measurements within the SM: 1. the weak
scale gauge couplings nearly unify under renormalization group evolution
at energy scale $m_{GUT}\simeq 2\times 10^{16}$ GeV in the MSSM, but not
the SM~\cite{gcu}, 2. the measured value of top quark mass falls within the
range needed to initiate a radiative breakdown of electroweak symmetry
in the MSSM~\cite{top}, 3. the measured value of the Higgs boson mass
$m_h\simeq 125$ GeV falls within the narrow range of MSSM predicted
values~\cite{mhiggs}, and 4. precision electroweak measurements actually
favor {\it heavy SUSY} over the SM~\cite{sven}.

Recent LHC searches with $\sqrt{s} = $ 13 TeV and integrated luminosity
$L = 139 \; {\rm fb}^{-1}$ have put lower bounds on the mass of the
gluino of about 2.2 TeV~\cite{atlasg,cmsg} and on the mass of top squark
of about 1.1 TeV~\cite{atlast,cmst}.  These limits, which have been
obtained using simplified model analyses assuming that the sparticle
spectrum is not compressed, fall well above upper bounds derived from
early naturalness considerations~\cite{bg,dg,ac,unn1,unn}.  However, the
naturalness estimates from the log-derivative measure are highly
dependent on what one regards as independent parameters of the
theory~\cite{dew}.\footnote{The various soft SUSY breaking terms which
  are adopted for the log-derivative measure are introduced to
  parameterize one's ignorance of how soft terms arise. In more
  UV-complete models such as string theory, then the various soft terms
  are all calculable and not independent. Ignoring this
  could result in an over-estimate of the UV sensitivity of the theory by
  orders of magnitude.}
We adopt the more conservative quantity $\Delta_{EW}$, that allows for
the possibility of correlations among model parameters, as a measure of
naturalness~\cite{rns}.  $\Delta_{EW}$ can be extracted from
Eq.~(\ref{eq:mzs}),
\begin{equation}
\frac{m_Z^2}{2} =\frac{m_{H_d}^2+\Sigma_d^d-(m_{H_u}^2+\Sigma_u^u)\tan^2\beta}{\tan^2\beta-1}-\mu^2\;,
\label{eq:mzs}
\end{equation}
which relates the mass of Standard Model $Z$ boson to SUSY Lagrangian
parameters at the weak scale and is obtained from the minimization conditions
of the MSSM scalar potential~\cite{baerbook}.
The electroweak fine-tuning parameter $\Delta_{EW}$ is defined by, 
\begin{equation}
\Delta_{EW}\equiv \mathrm(\rm {max |term\ on\ RHS\ of\ Eq.~\ref{eq:mzs}}|)/(m_Z^2/2) .
\label{eq:ew}
\end{equation}

The condition for naturalness is that the maximal contribution to the $Z$ mass
should be within a factor of several of its measured value. We consider
spectra that yield $\Delta_{EW}>\Delta_{EW}(\rm max) = 30$ as fine-tuned~\cite{upper}. 

This condition then requires : 
\begin{itemize}
    \item the SUSY-conserving $\mu$ parameter $\approx$ 110-350 GeV;

    \item the up-Higgs soft mass term $m_{H_u}^2$ may be large at high
      scales but can be radiatively-driven to (negative) natural values
      $\sim -m_{weak}^2$ at the weak scale;

    \item The finite radiative correction $\Sigma_u^u(\tilde{t}_{1,2})$
      has an upper bound of $(350$ GeV)$^2$ which is possible even for
      $m_{\tilde{t}}$ up to 3 TeV and $m_{\tilde{g}}$ $\approx$ 6
      TeV~\cite{bounds}, compatible with LHC constraints;

    \item the heavy Higgs masses $m_{A,H,H^\pm}\sim |m_{H_d}|$, with
     $|m_{H_d}|/\tan\beta\sim {m_Z\over \sqrt{2}} \sqrt{\Delta_{EW}}$. 
\end{itemize}
We thus see that naturalness requires \cite{bbbms} 
\be 
m_A\alt  \frac{m_Z\tan\beta\sqrt{\Delta_{EW}(\rm max)}} {\sqrt{2}},  
\ee 
and further, that for
$\tan\beta\sim 5-50$, the heavy Higgs boson masses may be expected to
lie in the (multi)-TeV range for an electroweak fine-tuning of up to a
part in thirty.  

The conditions mentioned above are satisfied in radiatively-driven
natural supersymmetric (RNS) models.  One of the features of RNS models
is that the heavier Higgs bosons may lie in the multi-TeV range while at
least some of the electroweakinos (EWinos) are below a few hundred
GeV. This means that generically we expect that in natural SUSY models
the supersymmetric decay modes of the heavy Higgs bosons should
be kinematically accessible, and often with branching
fractions comparable to SM decay modes.  If SUSY decay modes of the
heavy Higgs bosons are allowed, then 1. SM search modes will be
suppressed due to the presence of the SUSY decay modes and
2. potentially new avenues for heavy Higgs discovery may open up.  This
situation was investigated long ago under the supposition that the
lightest EWinos were predominantly gaugino-like~\cite{bbdkt}.
In Ref.~\cite{bbkt}, a lucrative $A,H\to\tchi_2^0\tchi^0_2\to
4\ell+\eslt$ search mode was identified for LHC.  However, in RNS
models, we expect instead that the lightest EWinos to be dominantly
higgsino-like.

\begin{figure}[h!]
\begin{center}
\includegraphics[scale=0.65]{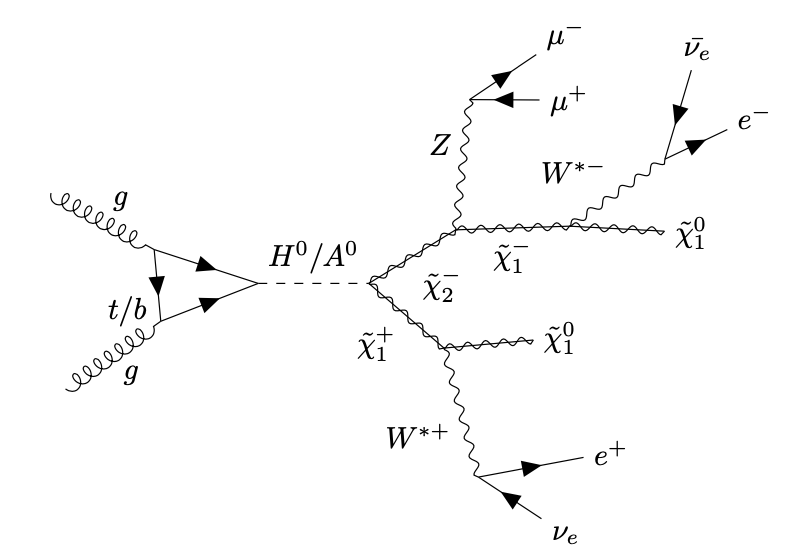}
\caption{Feynman diagram for
 $gg \to H,\ A (\to \tchi_2^\pm \tchi_1^\mp \to 4\ell
  +\slashed{E}_T)+X$ production; there is a similar diagram for
$H,\ A$ production via $b\bar{b}$ fusion.}
\label{fig:diagram}
\end{center}
\end{figure}

Thus, we explore here a new possible heavy Higgs discovery channel for
SUSY models with light higgsinos.  We identify the dominant new SUSY
decay mode for heavy neutral Higgs in natural SUSY models as
$H,\ A\to \tchi_1^\pm\tchi_2^\mp$ that proceeds with full gauge
strength\footnote{By full gauge strength, we only mean that the Higgs
  scalar-higgsino-gaugino vertex is unsuppressed. We recognize, of
  course, that the overall coupling of the heavy Higgs sector to the
  gauge boson sector is suppressed by mixing angles in the scalar Higgs
  sector.} (provided that the decay is kinematically
allowed). Allowing for chargino cascade decays, then an analogous
clean $4\ell +\eslt$ signature can be found.  It includes leptons from
the lighter chargino decay $\tchi_1^-\to\ell^-\bar{\nu}_\ell\tchi_1^0$
where the final state leptons are expected to be quite soft in the
chargino rest frame due to the expected small mass gap
$m_{\tchi_1^-}-m_{\tchi_1^0}$.  However, due to $m_{H,A}$ lying in the
TeV-range, these final state leptons may be strongly boosted and thus
can potentially contribute to the signal.  In this paper, we examine
the particular reaction $pp\to H,\ A\to \tchi_1^\pm\tchi_2^\mp\to
4\ell +\eslt$ where due to the heavy Higgs resonance, we expect
$M_T(4\ell,\eslt )$ to be kinematically bounded by $m_{H,A}$ (see
Fig.~\ref{fig:diagram}).  While this reaction will prove
difficult to extract at HL-LHC -- due in part to the several leptonic
branching fractions which are required -- we find that discovery in
this channel should be possible at the FCC-hh\cite{Bordry:2018gri} or
SPPC\cite{CEPC-SPPCStudyGroup:2015csa} $pp$ collider with
$\sqrt{s}\sim 100$ TeV and 15 ab$^{-1}$ of integrated luminosity.  The
FCC-hh or SPPC collider has emerged as the next target hadron collider
for CERN after HL-LHC in the updated European strategy
report~\cite{euro}.

To be specific, we will adopt a RNS benchmark (BM) point as listed in
Table \ref{tab:bm}, as generated using Isajet 7.88~\cite{isajet}.  This
BM comes from the two-extra-parameter non-universal Higgs model
NUHM2~\cite{nuhm2}. The NUHM2 model parameter space is given by
$m_0,m_{1/2},A_0,\tan\beta$ along with non-universal Higgs mass soft
terms $m_{H_u}\ne m_{H_d}\ne m_0$.  Using the EW minimization
conditions, it is convenient to trade the high scale soft terms
$m_{H_u},\ m_{H_d}$ for the weak scale parameters $\mu$ and $m_A$.  This
BM point yields $m_{\tg}\simeq 2.4$~TeV, somewhat beyond the LHC lower
limit of 2.2~TeV obtained from a {\em simplified model analysis}.
The heavy neutral Higgs scalars have mass $m_{H,A}\sim 1.2$ TeV which is
somewhat beyond the recent ATLAS limit\cite{ATLAS-H} that
requires $m_{H,A}\agt 1$ TeV
for $\tan\beta =10$ via an $H,\ A\to\tau^+\tau^-$ search at $\sqrt{s}=13$ TeV
and 139 fb$^{-1}$ of
integrated luminosity (while assuming no SUSY decay modes of the heavy Higgs
bosons).
Also, the SUSY $\mu$ parameter is taken to be $\mu =200$ GeV so that the
BM point lies just beyond the recent analyses of the {\em soft dilepton}
plus monojet higgsino signal\cite{lhcsoftdilep}.
For the listed BM point, the lighter EWinos $\tchi_{1,2}^0$ and $\tchi_1^\pm$
are higgsino-like while $\tchi_3^0$ is bino-like and $\tchi_4^0$
and $\tchi_2^\pm$ are wino-like.
\begin{table}[h!]
\centering
\begin{tabular}{lc}
\hline
parameter & NUHM2 \\
\hline
$m_0$      & 5 TeV \\
$m_{1/2}$      & 1.0 TeV \\
$A_0$      & -8.3 TeV \\
$\tan\beta$    & 10  \\
\hline
$\mu$          & 200 GeV  \\
$m_A$          & 1.2 TeV \\
\hline
$m_{\tilde{g}}$   & 2423 GeV \\
$m_{\tilde{u}_L}$ & 5293 GeV \\
$m_{\tilde{u}_R}$ & 5439 GeV \\
$m_{\tilde{e}_R}$ & 4804 GeV \\
$m_{\tilde{t}_1}$& 1388 GeV \\
$m_{\tilde{t}_2}$& 3722 GeV \\
$m_{\tilde{b}_1}$ & 3756 GeV \\
$m_{\tilde{b}_2}$ & 5150 GeV \\
$m_{\tilde{\tau}_1}$ & 4727 GeV \\
$m_{\tilde{\tau}_2}$ & 5097 GeV \\
$m_{\tilde{\nu}_{\tau}}$ & 5094 GeV \\
$m_{\tilde{\chi}_1^\pm}$ & 208.4 GeV \\
$m_{\tilde{\chi}_2^\pm}$ & 856.7 GeV \\
$m_{\tilde{\chi}_1^0}$ & 195.4 GeV \\ 
$m_{\tilde{\chi}_2^0}$ & 208.5 GeV \\ 
$m_{\tilde{\chi}_3^0}$ & 451.7 GeV \\ 
$m_{\tilde{\chi}_4^0}$ & 867.9 GeV \\ 
$m_h$       & 125.0 GeV \\ 
\hline
$\Omega_{\tilde{z}_1}^{std}h^2$ & 0.011 \\
$BF(b\to s\gamma)\times 10^4$ & $3.2$ \\
$BF(B_s\to \mu^+\mu^-)\times 10^9$ & $3.8$ \\
$\sigma^{SI}(\tilde{\chi}_1^0, p)$ (pb) & $3.1\times 10^{-9}$ \\
$\sigma^{SD}(\tilde{\chi}_1^0, p)$ (pb)  & $6.1\times 10^{-5}$ \\
$\langle\sigma v\rangle |_{v\to 0}$  (cm$^3$/sec)  & $2.0\times 10^{-25}$ \\
$\Delta_{\rm EW}$ & 25.5 \\
\hline
\end{tabular}
\caption{Input parameters (TeV) and masses (GeV)
for a SUSY benchmark point from the NUHM2 model
with $m_t=173.2$ GeV using Isajet 7.88~\cite{isajet}.
}
\label{tab:bm}
\end{table}

\subsection{Review of some previous related work and plan for this work}
\label{ssec:review}

SUSY Higgs boson decays to EWinos were first calculated in Baer
{\it et al.} Ref.~\cite{bddt}.  A more comprehensive treatment was given
in Gunion {\it et al.} \cite{ssc1} and Gunion and Haber
\cite{gh3}. Griest and Haber \cite{ghaber} considered the effect of
invisible Higgs decays $H\rightarrow \tz_1\tz_1$.  In Kunszt and Zwirner
Ref.~\cite{kz}, the phenomenology of SUSY Higgs bosons in the $m_A$
vs. $\tan\beta$ plane with just SM decay modes was considered in light
of the important radiative corrections to $m_h$.  The $m_A$
vs. $\tan\beta$ plane was mapped including the effects of Higgs to SUSY
decays in Baer {\it et al.} Ref.~\cite{bbdkt} where diminution of SM
Higgs decay channels due to SUSY modes was considered along with the
potential for new discovery channels arising from the SUSY decay
modes. In Ref.~\cite{bbkt}, the discovery channel $H,\ A\to
\tz_2\tz_2\to 4\ell +\eslt$ was examined.  In Djouadi {\it et al.}
\cite{djkz}, SUSY decays of heavy Higgs bosons at $e^+e^-$ colliders were
considered.  Barger {\it et al.} in Ref.~\cite{bbgh} examined
$s$-channel production of SM and SUSY Higgs bosons at muon colliders.
In Belanger {\it et al.} \cite{belanger}, SUSY decays of Higgs bosons at
LHC were examined. Choi {\it et al.}\cite{cdls} examined the effects of
CP violating phases on Higgs to SUSY decays.  In Ref.~\cite{css}, a CMS
study of $H,\ A\to \tz_2\tz_2\to 4\ell +\eslt$ was performed.
In Ref. \cite{bisset}, signals from $H,\ A\to 4\ell+\eslt$ were examined
including {\it all} SUSY cascade decays of heavy Higgs bosons in
scenarios where the $\tchi_1^0$ was bino-like.
In Bae {\it et al.}\cite{bbbms}, the impact of natural SUSY with light
higgsinos on SUSY Higgs phenomenology was examined and natural regions
of the $m_A$ vs. $\tan\beta$ plane were displayed along with relevant
SUSY Higgs branching fractions.  The LHC SUSY Higgs signatures $H,
A\to mono-X+\eslt$ (where $X=W,\ Z,\ h$) were examined against huge SM
backgrounds.  In Bae {\it et al.} Ref.~\cite{bbns}, the effect of
natural SUSY on Higgs coupling measurements $\kappa_i$ was examined.  In
Barman {\it et al.}~\cite{bbcc}, SUSY Higgs branching fractions and
$mono-X+\eslt$ signatures were examined at LHC for several benchmark
points along with a Higgs to SUSY trilepton signature.
In Ref. \cite{bagnaschi}, six MSSM SUSY Higgs benchmark points were proposed
for LHC search studies, including one with a low, natural value
of $\mu$ (which seems now to be LHC-excluded).
Gori, Liu and Shakya examined SUSY Higgs decays to EWinos and to stau pairs in
Ref.~\cite{gls}. In Adhikary {\it et al.}~\cite{abgkk}, Higgs decay to
EWinos at LHC were examined, especially the $Z+\eslt$ and $h+\eslt$
signatures along with the possibility of Higgs decays to long-lived
charged particles (LLCPs).

\subsection{Plan for this paper}
\label{ssec:plan}

In the present paper, we examine Higgs decays to SUSY particles in
natural SUSY models with light higgsinos. In particular, in light of the
large SM backgrounds for $mono-X+\eslt$ searches, we examine the
viability of resurrecting the $H,\ A\to 4\ell+\eslt$ signature. In the
natural SUSY case, this signature could arise from
$H,\ A\to\tchi_1^\pm\tchi_2^\mp$ followed by $\tchi_2^\pm\to
Z\tchi_1^\pm$. The $Z\to \ell^+\ell^-$ decay should be easily visible
but the leptons from $\tchi_1^-\to\ell\nu_\ell\tchi_1^0$ are typically
very soft in the $\tchi_1^\pm$ rest frame. Owing to the TeV scale values
of $m_{H,A}$, these otherwise soft leptons may be boosted to detectable
levels. While such a complicated decay channel appears intractable at
HL-LHC, the FCC-hh or SPPC operating at $\sqrt{s}\sim 100$ TeV and 15
ab$^{-1}$ should allow for discovery for $m_{H,A}\sim 1-2$ TeV with
advanced machine learning (ML) techniques; here we have used boosted
decision trees as an illustration.\footnote{Since one of our goals is to
  illustrate how ML techniques may help to eke out a signal that lies
  below the discovery limit using standard cut-and-count analyses if the
  Higgs boson is very massive, we have confined our study to the signal
  in this single channel, and for simplicity carried out our
  calculations using parton level simulations.}

The remainder of this paper is organized as follows. In Sec. \ref{sec:prod},
we present $s$-channel production rates for heavy Higgs bosons at LHC14 and
at FCC-hh or SPPC. In Sec. \ref{sec:BFs}, we discuss the heavy Higgs branching 
fractions that are expected in natural SUSY models and we motivate our
particular four lepton SUSY Higgs discovery channel. 
In Sec. \ref{sec:BGs}, we discuss leading SM backgrounds to the 
$H,\ A\to 4\ell+\eslt$ signal channel.
In Sec. \ref{sec:cuts}, we perform a cut-based analysis while in 
Sec. \ref{sec:BDT} we show one can do much better
by invoking a boosted-decision-tree (BDT) analysis. In Sec. \ref{sec:conclude},
we summarize our main conclusions.

\section{Heavy Higgs production at LHC and FCC-hh or SPPC}
\label{sec:prod}

Here, we will focus on the $s$-channel heavy neutral Higgs boson
production reactions $pp\rightarrow H,\ A$ which occurs via the
gluon-gluon and $b\bar{b}$ fusion subprocesses.
Other reactions such as $pp\rightarrow qqH$ (VV fusion reactions)
$WH$, $ZH$ and $t\bar{t}H$ all occur at
lower rates~\cite{higgspro} 
and also lead to
different final state topologies.
Hence, we will not include these in our analysis.

\begin{figure}[h!]
  \centering
  \includegraphics[width=0.48\textwidth]{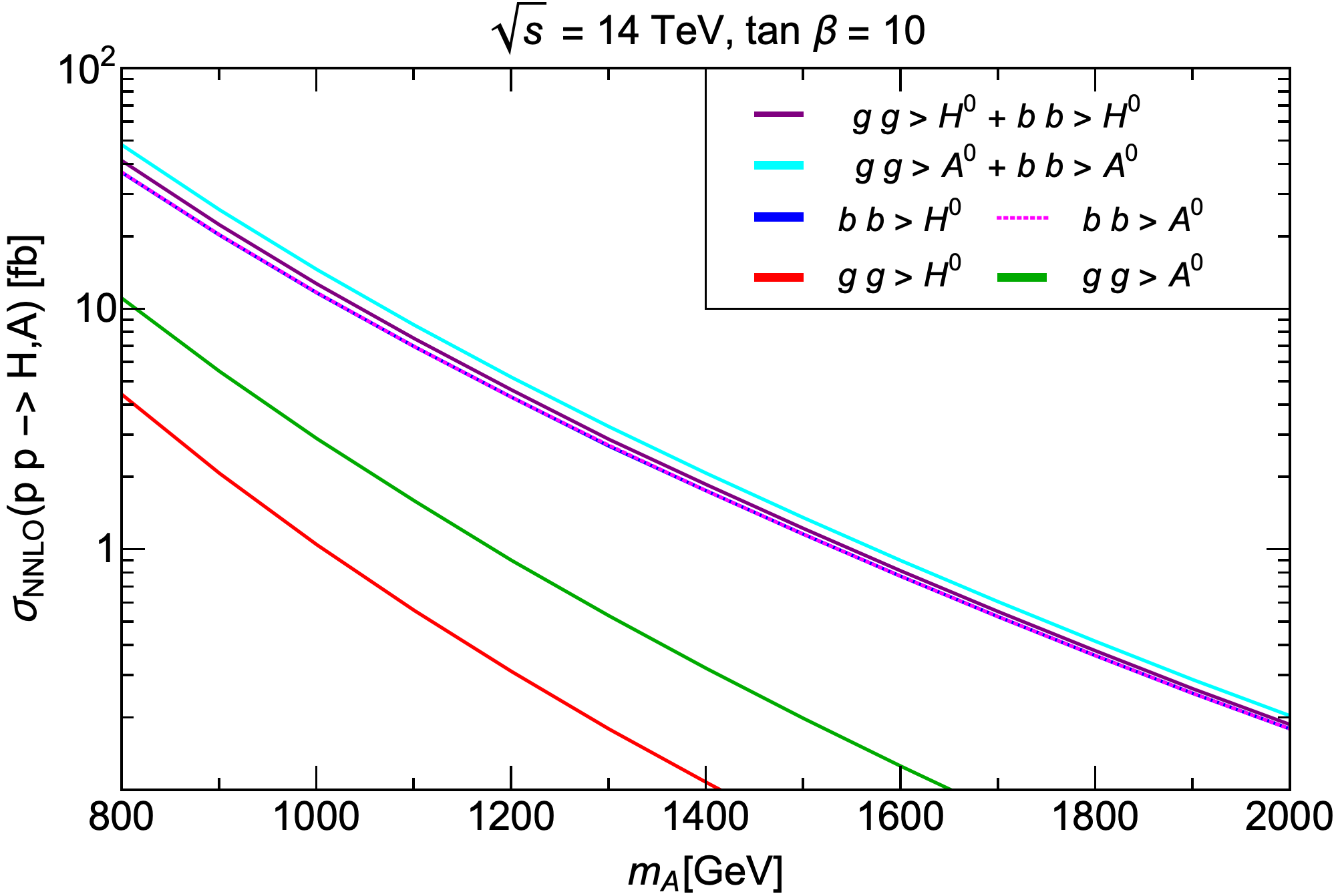}
   \hspace{1mm}
  \includegraphics[width=0.48\textwidth]{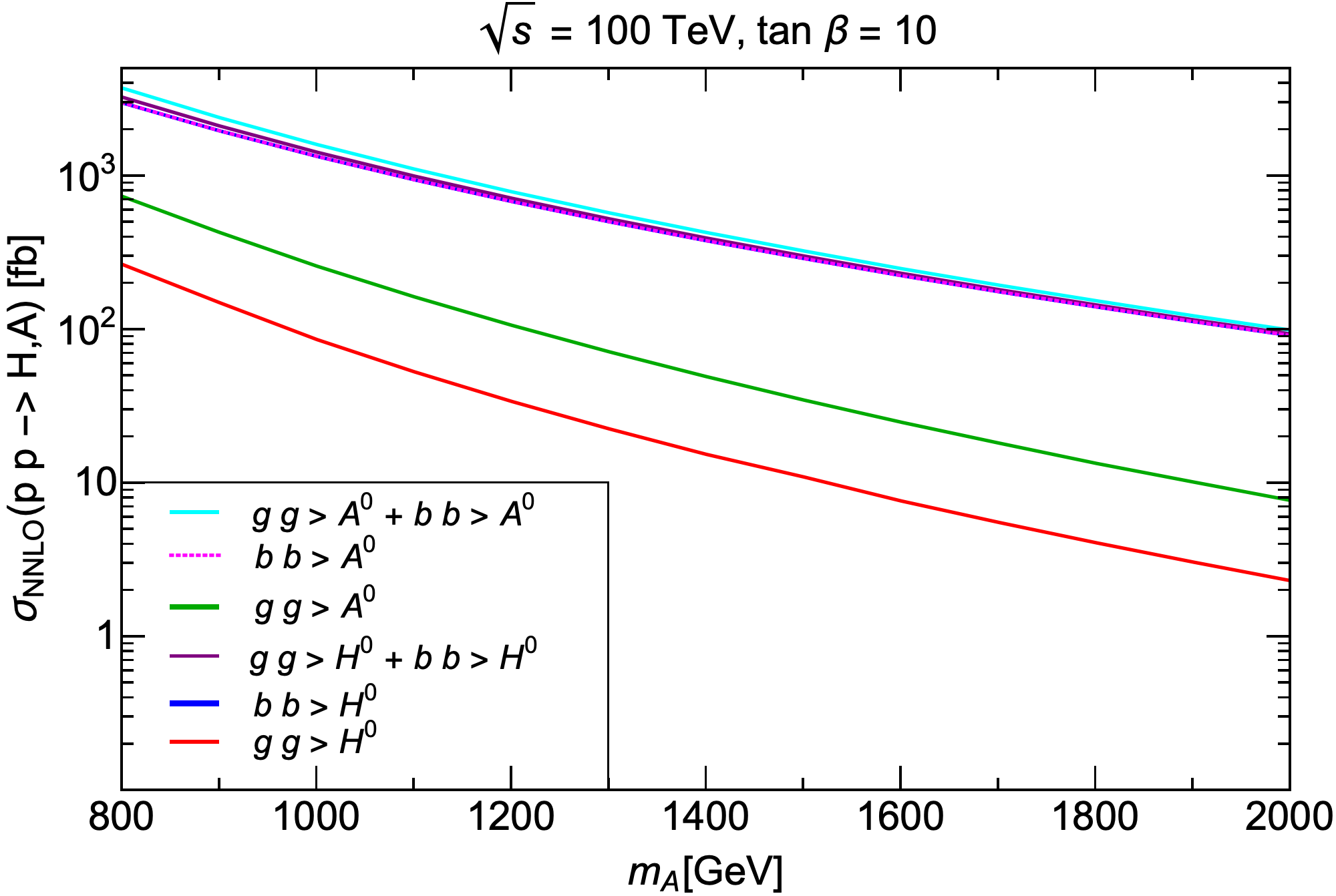}
  \caption{$\sigma_{NNLO}(pp\rightarrow H,\ A +X)$ 
    for $s$-channel heavy neutral Higgs boson production reactions
    via $gg$ and $b\bar{b}$ fusion versus
$m_A$ for (a) $\sqrt{s} =$ 14 and (b) $\sqrt{s} =$ 100 TeV.
We take $\tan\beta =10$.
Results are from SusHi~\cite{Harlander:2012pb,Harlander:2016hcx}.
}  
\label{fig:sigma}
\end{figure}

In Fig.~\ref{fig:sigma}, we show the heavy neutral Higgs production
cross sections at next-to-next-to-leading order (NNLO) in QCD.
We adopt the SusHi program~\cite{
Harlander:2012pb,Harlander:2016hcx,Harlander:2002wh,Harlander:2003ai,Aglietti:2004nj,Bonciani:2010ms,Degrassi:2010eu,Degrassi:2011vq,Degrassi:2012vt,Harlander:2005rq,Chetyrkin:2000yt}
to generate these results, which include QCD corrections and effects
from top and bottom squark loops.
Higher order QCD corrections typically boost these cross sections above
their leading order estimates.  Frame (a) shows results for
$\sqrt{s}=14$ TeV while frame (b) shows results for $\sqrt{s}=100$ TeV.
We see that even for $\tan\beta=10$, heavy Higgs boson production via
$b\bar{b}$ fusion dominates that from gluon fusion.  From frame (a), we
see that for $m_A\sim 800$ GeV, the total production cross sections
occur for both $H$ and $A$ production at the $\sim 40$ fb level. As
$m_A$ increases, the rates fall and are already below the 0.2 fb level
for $m_A\agt 2$ TeV.  We can anticipate that once we fold in various
leptonic branching fractions and include detector acceptances, we will
not expect very high rates for multi-lepton signals from heavy neutral
SUSY Higgs bosons at LHC14. In frame (b), we show the results for
$\sqrt{s}=100$ TeV. Here, the cross sections are increased by factors of
70-500 as $m_A$ varies from 800-2000 GeV.
%

%

\section{Heavy Higgs and sparticle branching fractions}
\label{sec:BFs}

\begin{figure}[h!]
  \centering
  \includegraphics[width=0.48\textwidth]{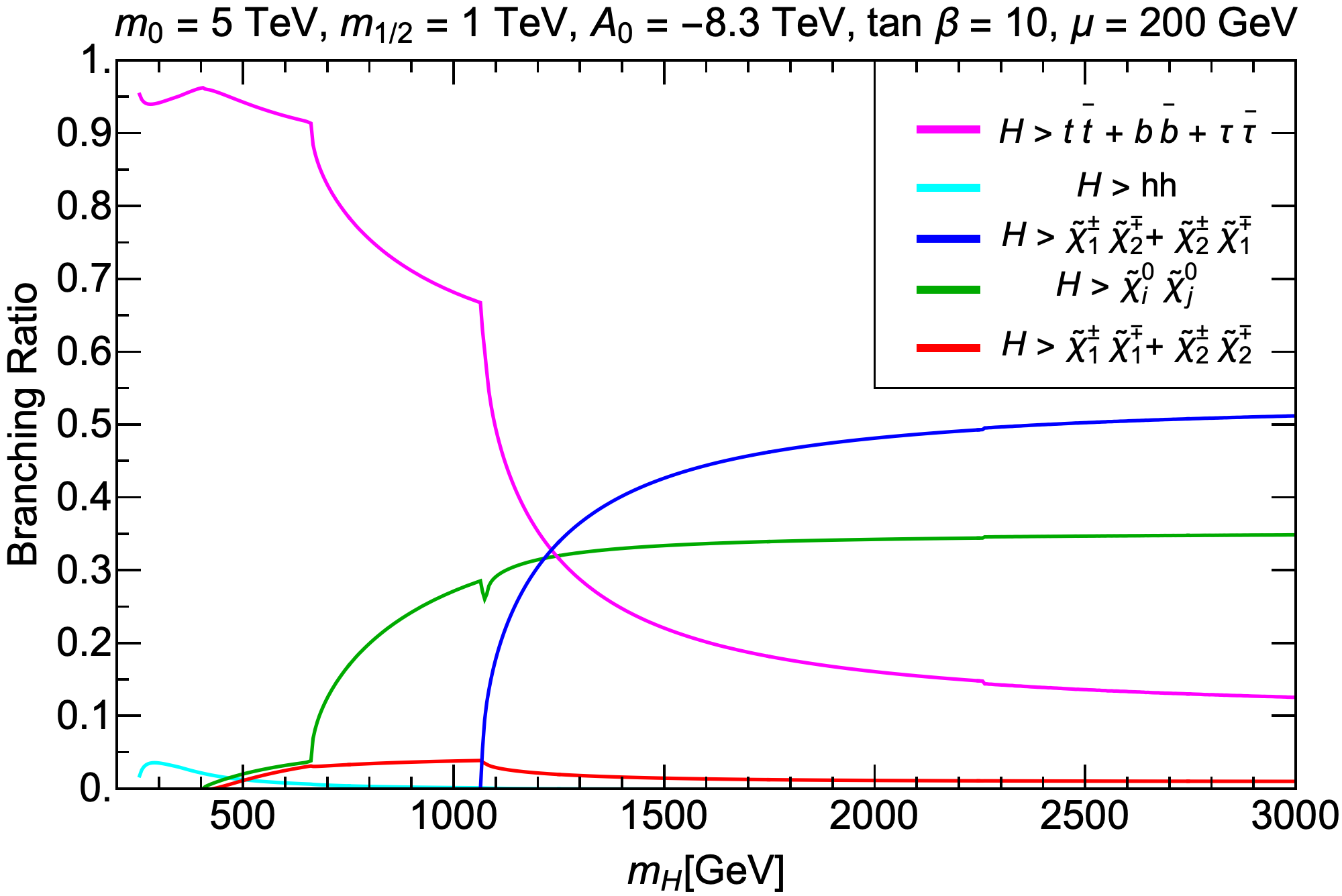} \\
  \includegraphics[width=0.48\textwidth]{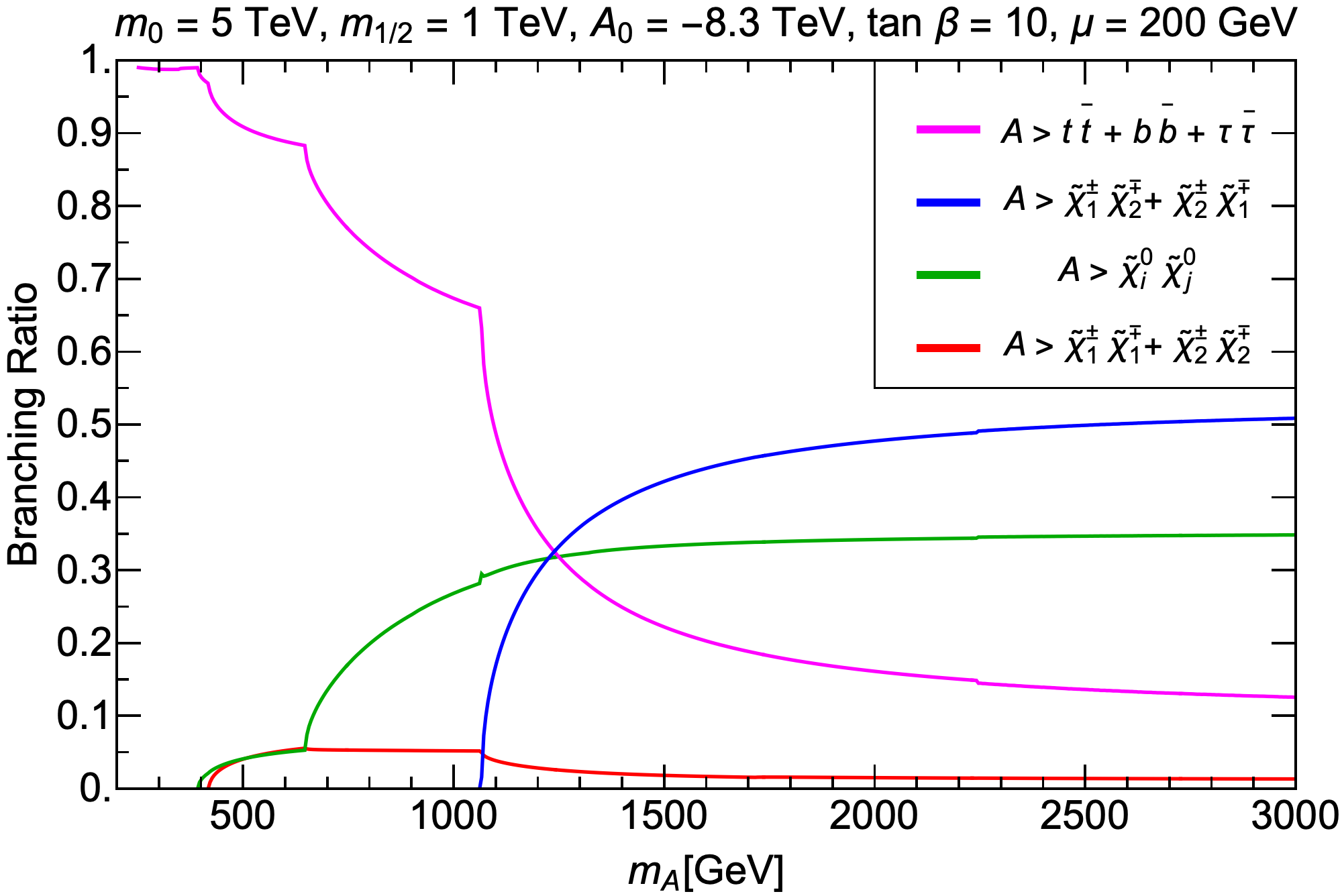}
   \hspace{1mm}
  \includegraphics[width=0.48\textwidth]{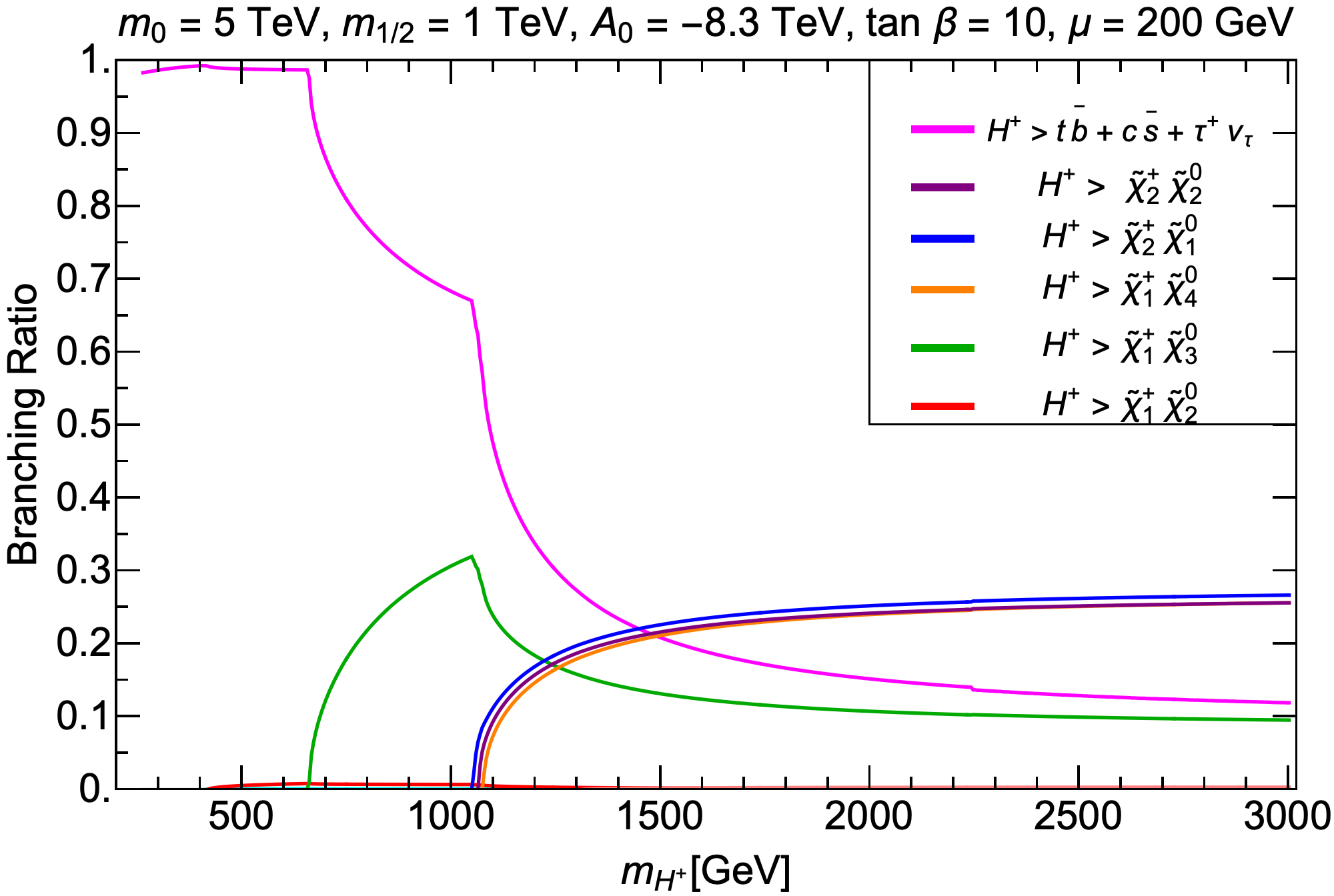}
  \caption{Branching fractions versus heavy Higgs mass for 
{\it a}) $H$, {\it b}) $A$ and {\it c}) $H^+$ into SM
and SUSY particles in the NUHM2 model with $\mu= 200$ GeV and
$m_0=5$ TeV, $m_{1/2}=1$ TeV, $A_0=-8.3$ TeV and $\tan\beta =10$.
}  
\label{fig:BFs}
\end{figure}

In this Section, we present some updated heavy neutral and charged
Higgs branching fractions which we extract from the Isajet 7.88
code~\cite{isajet}.  We adopt the benchmark point from Table
\ref{tab:bm} except now we allow the heavy Higgs mass $m_A$ to vary.
In frame (a), we show branching fractions for
the heavy neutral scalar $H$.  At low $m_H$, the SM modes
$H\rightarrow b\bar{b}$, $\tau\bar{\tau}$ and $t\bar{t}$ are dominant,
with their exact values depending on $\tan\beta$ (large $\tan\beta$
enhances the $b\bar{b}$ and $\tau\bar{\tau}$ modes). For $m_H\sim
400-650$ GeV, the SM modes are still dominant even though the light
electroweakino modes are open. We can understand this by examining the
Higgs sector Lagrangian in the notation of Ref.~\cite{baerbook},
Sec. 8.4: \be {\cal L}\ni -\sqrt{2}\sum_{i,A}{\cal S}_i^\dagger
gt_A\bar{\lambda}_A\frac{1-\gamma_5}{2} \psi_i +h.c.  \ee
where ${\cal S}_i$ labels various matter and Higgs scalars (labeled by
$i$), $\psi_i$ is the fermionic superpartner of ${\cal S}_i$, and
$\lambda_A$ is the gaugino with gauge index $A$.  Also, $g$ is the gauge
coupling for the gauge group and $t_A$ are the corresponding gauge group
generator matrices.  Letting ${\cal S}_i$ be the Higgs scalar fields,
then we see that the Higgs-EWino coupling is maximal when there is
little mixing in that the Higgs fields couple directly to gaugino plus
higgsino.  Back in Fig. \ref{fig:BFs}(a), for $m_H$ small, then the
only open decay modes are $H$ to higgsino plus higgsino, and so the
coupling must be dynamically suppressed because the gaugino component of
the lightest EWinos is very small. Thus the SM modes are still
dominant. As $m_H$ increases, then the decay to gaugino plus higgsino
turns on and the above coupling is unsuppressed (as has also been noted
in footnote 2, above).  For our choice of SUSY parameters, this happens
around $m_H\sim 650$ GeV for $H$ decay to higgsino plus bino and around
$m_H\sim 1050$ GeV for $H$ decay to wino plus higgsino. Since the latter
coupling involves the larger $SU(2)_L$ gauge coupling, the decay
$H\rightarrow$ wino plus higgsino ultimately dominates the branching
fraction once it is kinematically allowed.  Thus, for $m_H\agt 1250$
GeV, $H\rightarrow \tchi_1^\pm\tchi_2^\mp$ dominates the branching
fraction (blue curve), while decays of $H$ to the lighter neutral
higgsino-like neutralino plus the heavier neutral wino or bino-like
neutralino (green curve) have a branching fraction about half as large.
In this range of $m_H$, the SM $H$ decay modes are severely depressed
from their two-Higgs doublet (non-SUSY) expectation. This will make
heavy Higgs detection via $t\bar{t}$, $b\bar{b}$ and $\tau\bar{\tau}$
much more difficult.  On the other hand, it opens up new discovery
channels by searching for the dominant $H\rightarrow$ EWino modes.

In Fig. \ref{fig:BFs}(b), we show the same branching fractions
except now for the pseudoscalar $A$. The branching fractions look
qualitatively similar to those in frame (a) since the same
reasoning applies. Thus, the $A$ will decay mainly to SM modes for 
smaller values of $m_A$ even though decays to higgsino-like pairs are
available. It is only when decays to gaugino plus higgsino open up that
the branching fractions to SUSY modes begin to dominate.

For completeness, we also show in Fig. \ref{fig:BFs}(c) the
branching fractions for charged Higgs decays $H^+$. As in the previous
cases, $H^+$ decay to SM modes $t\bar{b}$ and $\tau^+\nu_\tau$ dominate
at low values of $m_{H^+}$ even though decay to $\tchi_1^+\tz_{1,2}$
modes are kinematically allowed.  As $m_{H^+}$ increases, then decays to
$\tchi_1^+\tz_3$ (higgsino-bino) followed by $\tchi_2^+\tz_{1,2}$
and $\tz_4\tchi_1^+$ (higgsino-wino) turn on and rapidly dominate the
decays.

Some dominant heavy neutral Higgs decay branching fractions are shown in
Table \ref{tab:HABF} for the benchmark point shown in Table
\ref{tab:bm}. We see again that for the benchmark point the $H,\ A$
decays to SM modes are suppressed compared to decay rates into gaugino
plus higgsino.
\begin{table}[h!]
\centering
\begin{tabular}{lc}
\hline
decay mode & BF \\
\hline
$H\to b\bar{b}$  & 22.5\% \\
$H\to\tchi_1^\pm\tchi_2^\mp$  & 31.2\% \\
$H\to\tchi_2^0\tchi_4^0$  & 12.2\% \\
\hline
$A\to b\bar{b}$  & 22.9\% \\
$A\to\tchi_1^\pm\tchi_2^\mp$  & 30.0\% \\
$A\to\tchi_1^0\tchi_4^0$  & 12.2\% \\
\hline
\end{tabular}
\caption{Dominant branching fractions for heavy Higgs $H,\ A$
for the benchmark point with $m_A=1200$ GeV.
}
\label{tab:HABF}
\end{table}

\begin{figure}[h!]
\begin{center}
\includegraphics[width=0.48\textwidth]{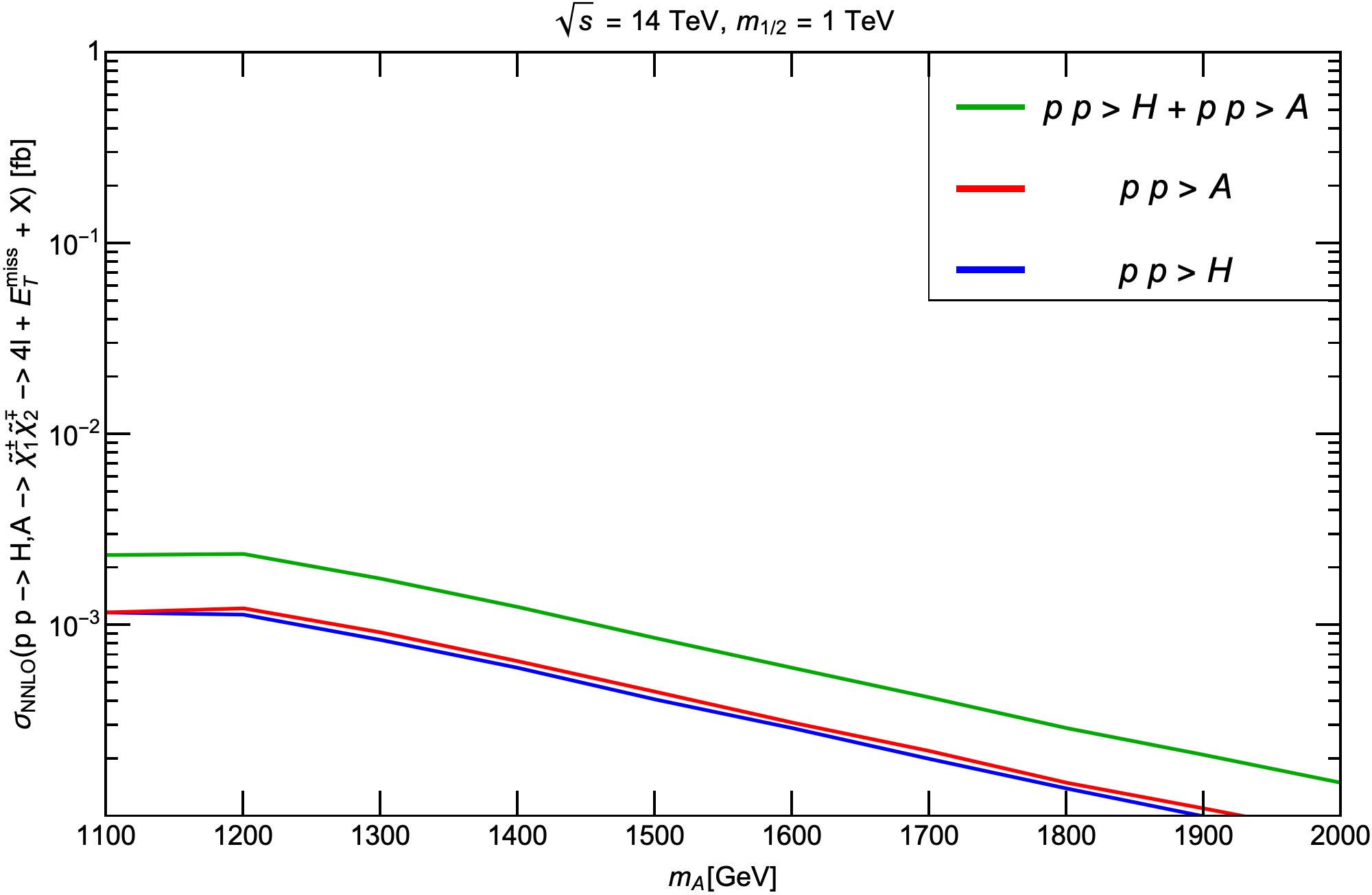}
\hspace{1mm}
\includegraphics[width=0.48\textwidth]{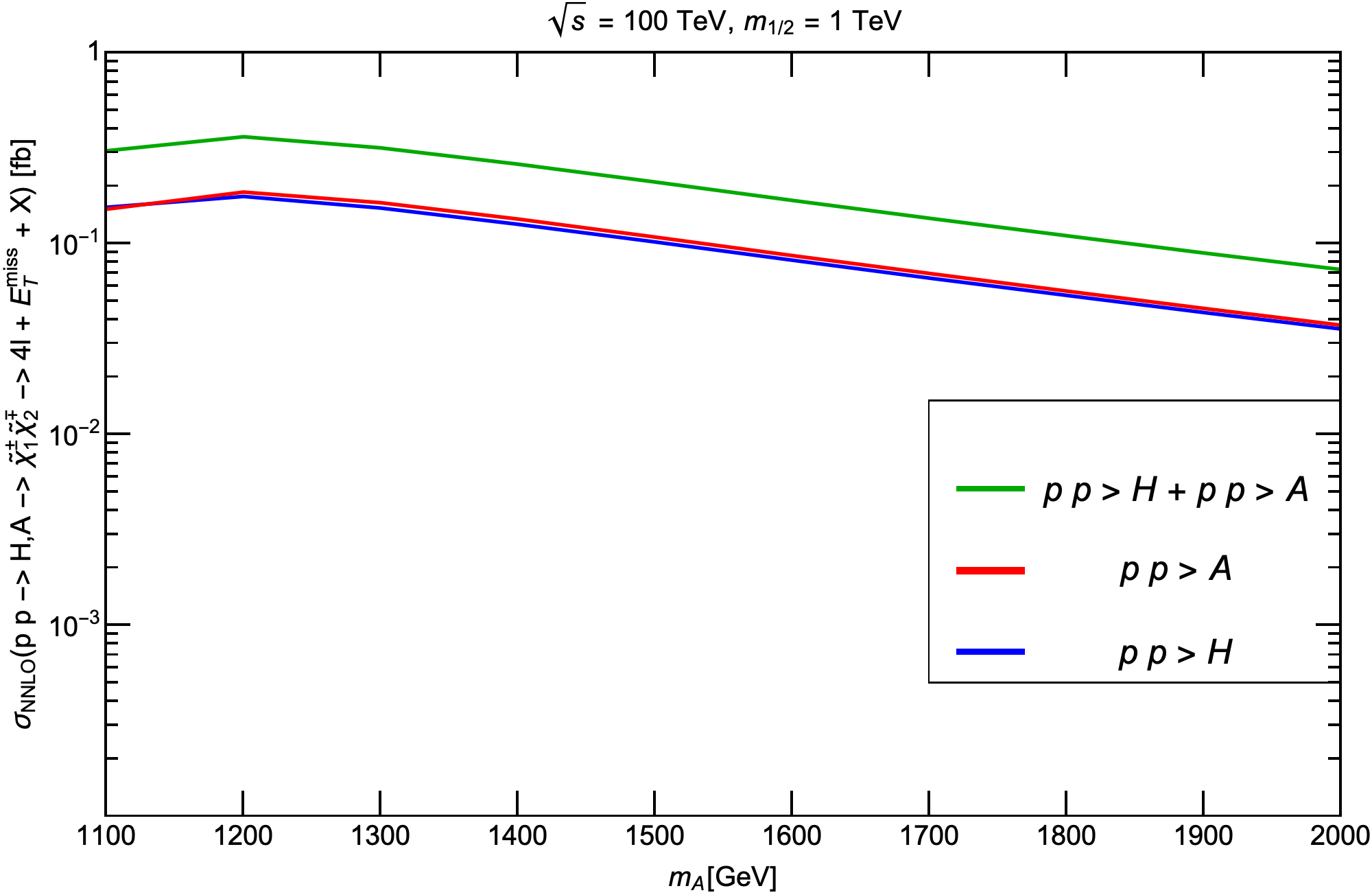}
\caption{NNLO Cross sections from SusHi $\sigma(A)$, $\sigma(H)$, and
  $\sigma(A)+\sigma(H)$ times the cascade decay branching fractions into
  the $4\ell +\eslt$ final state in fb vs. $m_A$ for (a) 14 TeV and (b)
  100 TeV without any cuts.}
\label{fig:sig4lMET}
\end{center}
\end{figure}

In Fig. \ref{fig:sig4lMET}, we combine the $H,\ A$ production rates from
Fig. \ref{fig:sigma} with the Higgs boson and sparticle branching
fractions to the $4\ell+\eslt$ final state depicted in
Fig. \ref{fig:diagram}. We see from Fig. \ref{fig:sig4lMET}(a) that, for
$\tan\beta=10$, even without cuts we expect {\em at most} $\sim 7$
signal events at HL-LHC, assuming an integrated luminosity of 3000
fb$^{-1}$.  Moreover, we expect that this will be reduced considerably
once detector efficiency and analysis cuts are folded in.  However, as
we can see from frame (b), the raw signal cross section is larger at the
higher energy FCC-hh or SPPC by a factor 150-500 (compared to LHC14), so
that with the projected 15~ab$^{-1}$ of integrated luminosity, we may
hope to be able to extract an observable signal even after cuts. We
will, therefore, mostly focus our attention on a 100~TeV $pp$ collider
in the remainder of this paper.

\begin{figure}[h!]
\begin{center}
\includegraphics[width=0.7\textwidth]{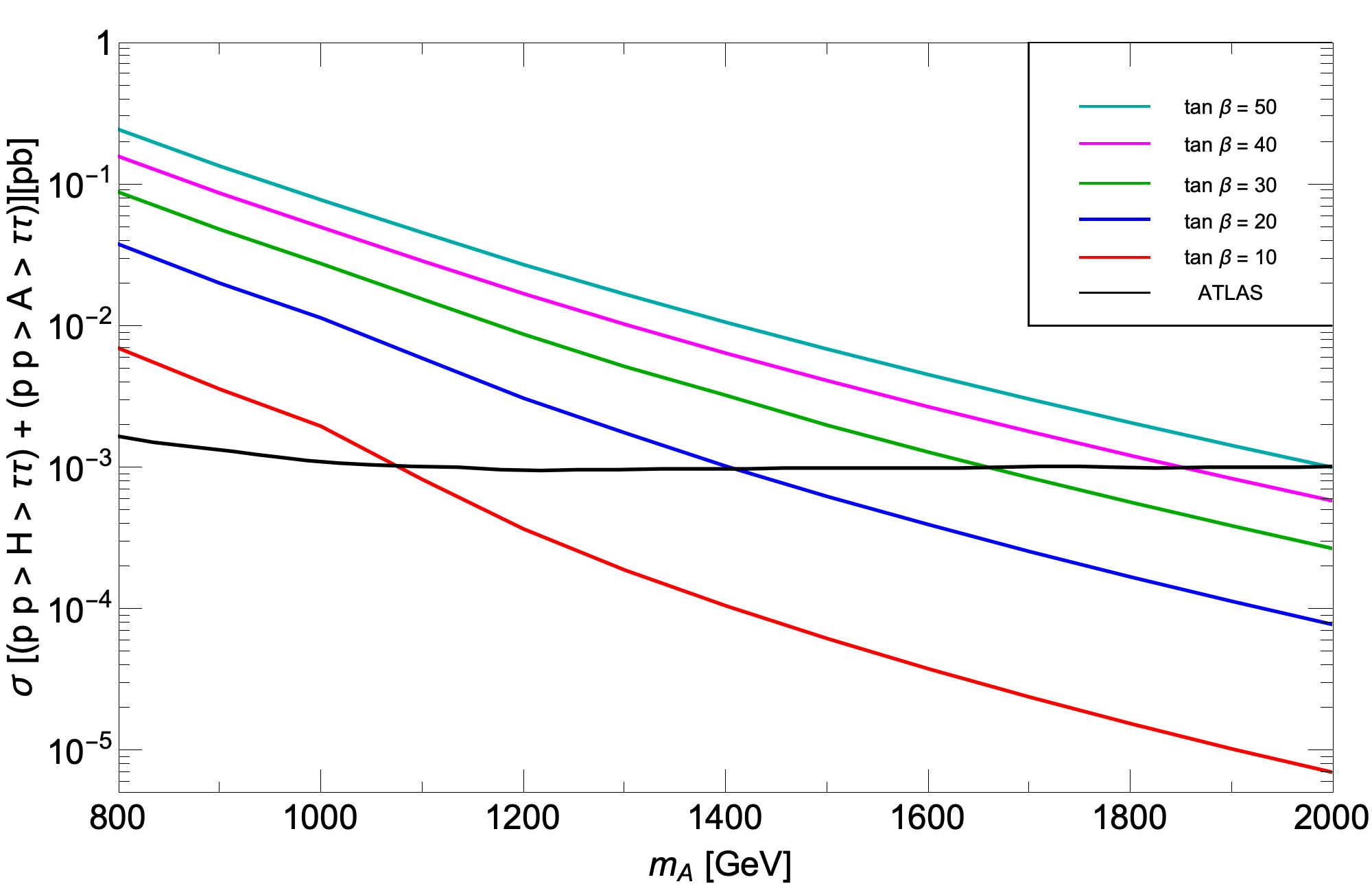}
\caption{The summed cross section times branching ratio for
  $A/H \to\tau\bar{\tau}$ versus $m_A$ at LHC14 for
  several values of $\tan\beta$. Other parameters are fixed at their
values for the model line introduced in the text. The horizontal black
line shows the current upper limit on the cross section obtained by ATLAS.}
\label{fig:HAlimit}
\end{center}
\end{figure}

The reader may be concerned that our dismissal of the possibility of a
signal in the $4\ell+\eslt$ channel at LHC14 was based on the event rate
for $\tan\beta=10$ when it is well-known that the couplings of the $A$
and $H$ both increase with $\tan\beta$, resulting in an increased rate
for $H/A$ production from bottom quark fusion. It should, however, be
remembered that the range of $m_A$ excluded by the current upper limit
on the cross section times branching ratio for the decay $\phi \to
\tau\bar{\tau}$ ($\phi = A,H$) also increases with $\tan\beta$ for this
same reason. This is illustrated in Fig.~\ref{fig:HAlimit} where we show
the expectations for the resonant production of tau pairs from the decay
of $H/A \to\tau\bar{\tau}$ versus $m_A$ for several values of
$tan\beta$. Other parameters are taken to be the same as for the
model-line introduced earlier. The horizontal black line is the current
ATLAS upper bound on this rate \cite{ATLAS-H}. We see that while $m_A >
1.1$~TeV for $\tan\beta=10$, for $\tan\beta=50$, $m_A > 2$~TeV. Scaling
the cross section in the left frame of Fig.~\ref{fig:sig4lMET} by the
ratio of the corresponding values of $\tan^2\beta$ still leaves us with
just a handful of events {\em before cuts} at the HL-LHC for currently
allowed values of $m_A$.

\section{SM Backgrounds and Analysis Cuts }
\label{sec:BGs}

Our signal $p p \to H,A \to \chi_1^{\pm}\chi_2^{\mp} \to 4 \ell +
\slashed{E}_T$ contains 4 leptons and missing energy in the final
states, where one pair of leptons comes from the decay of a
$Z$-boson. Since, as just mentioned, the signal rate is too small
at the HL-LHC, we will from now on mostly focus
our attention on a 100~TeV $pp$ collider.

Our simplified study has been carried out at parton level.
The dominant SM background to the $4\ell+\slashed{E}_T$ events comes
from $W^{\pm}W^{\mp}V$, $t\bar{t}V$, $Zh$ and $ZZV$ ($V = W^{\pm},Z,
\gamma$).  Notice that the partonic final states from
the signal, as well as from all the backgrounds other than
$t\bar{t}V$ production, are free of any hadronic activity.  We use
tree-level matrix elements from the HELAS library in Madgraph to
evaluate the backgrounds, and then scale our cross section to NLO with
$K$-Factors calculated using MCFM ~\cite{mcfm}.\footnote{The $K$-factors
  that we use are, $K_{WWV}= 1.36$, $K_{t\bar{t}V}=1.30$, $K_{Zh}=1.40$
  and $K_{ZZV}=1.40$. }  For the $t\bar{t} V$
background we veto events which contain any $b$-jets ({\it i.e. }
$b$-quarks) with $p_T > 20$ GeV and $|\eta(b)| < 2.5$.  This serves as a
powerful cut in reducing this background.  However, with PDF
enhancements, we find that this background becomes the second most
dominant background at $\sqrt{s} =$ 100~TeV. $W^{\pm}W^{\mp}V$ proves to
be the most dominant background at all energies.

To select events, we identify the isolated leptons if they satisfy 

\begin{itemize}
    \item $p_T$ ($\ell_1$, $\ell_2$, $\ell_3$, $\ell_4$) $>$ 20 GeV, 10
      GeV, 10 GeV, 10 GeV;

    \item $|\eta|$ ($\ell_1$, $\ell_2$, $\ell_3$, $\ell_4$) $<$ 2.5.
\end{itemize}
We model experimental errors in the measurement of lepton energies
by Gaussian smearing electron and muon energies using~\cite{ATLAS:2013-004}, 
  \begin{equation}
    \frac{\Delta E}{E} = \frac{0.25}{\sqrt{\rm E(GeV)}}\oplus 0.01,
  \end{equation}
where $\oplus$ denotes addition in quadrature.

Since the signal of interest has a final state of $4\ell +\slashed{E}_T$, 
we started with a set of minimal cuts, labeled as {\bf Cuts A}, which include : 

\begin{itemize}
    \item Veto events with $b$-jets $p_T$ (jet) $>$ 20 GeV and $|\eta|$
      (jet) $<$ 2.5 as already mentioned;

    \item $\Delta_R(j, \ell)>$ 0.4, where $j$ denotes a $b$-quark
      with $p_T < 20$~GeV or with $|\eta_b|>0.4$, to mimic lepton isolation;

    \item Invariant mass for two opposite sign same flavor leptons
      $M_{\ell^+\ell^-}$ $>$ 10 GeV, to reduce the background from
      $\gamma^*\to\ell\bar{\ell}$;
    
\item $\slashed{E}_T > 125$ GeV.
\end{itemize}

\begin{figure}[h!]
  \centering
  {\includegraphics[width=0.48\textwidth]{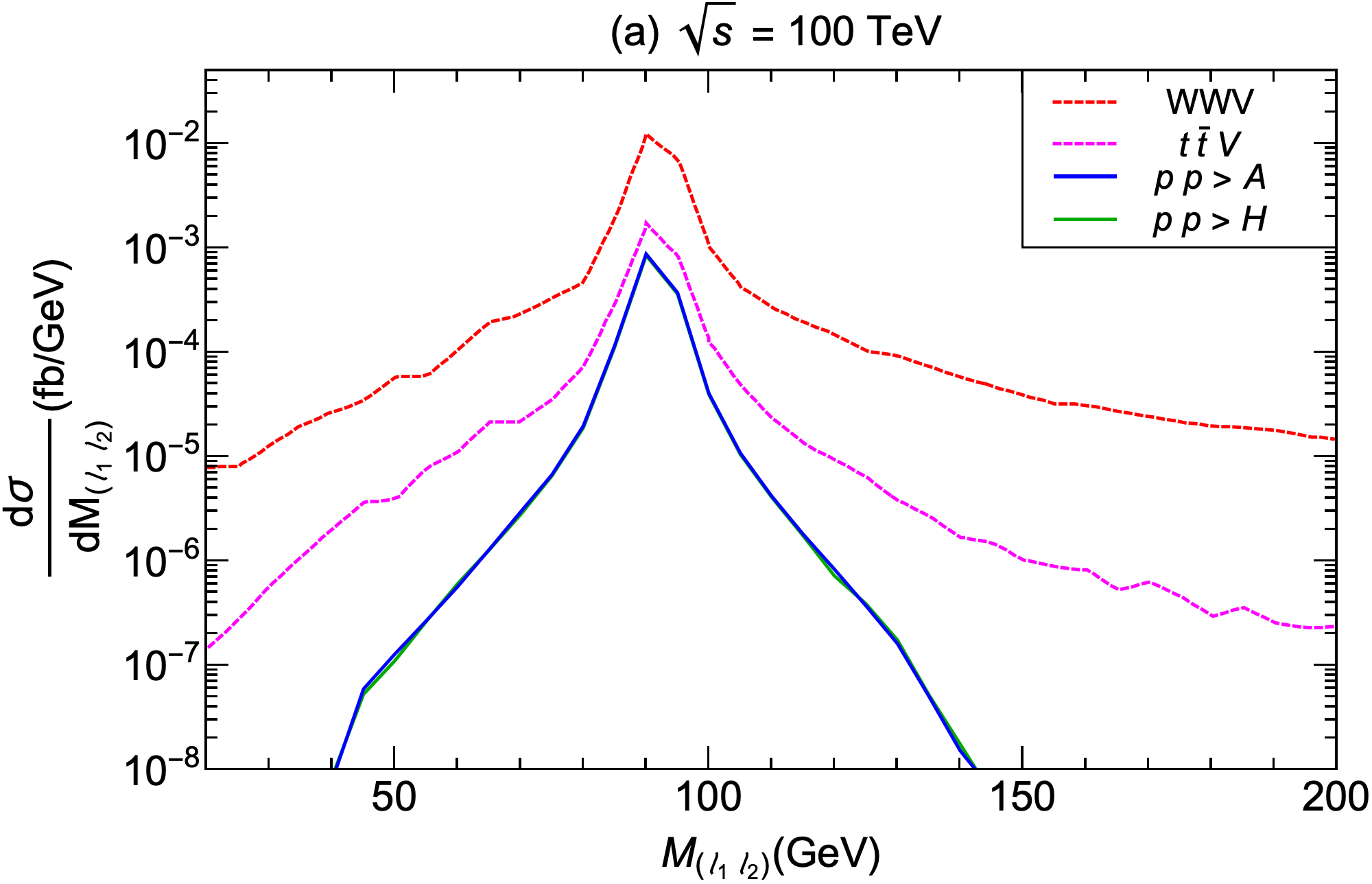}}
   \hspace{1mm}
  {\includegraphics[width=0.48\textwidth]{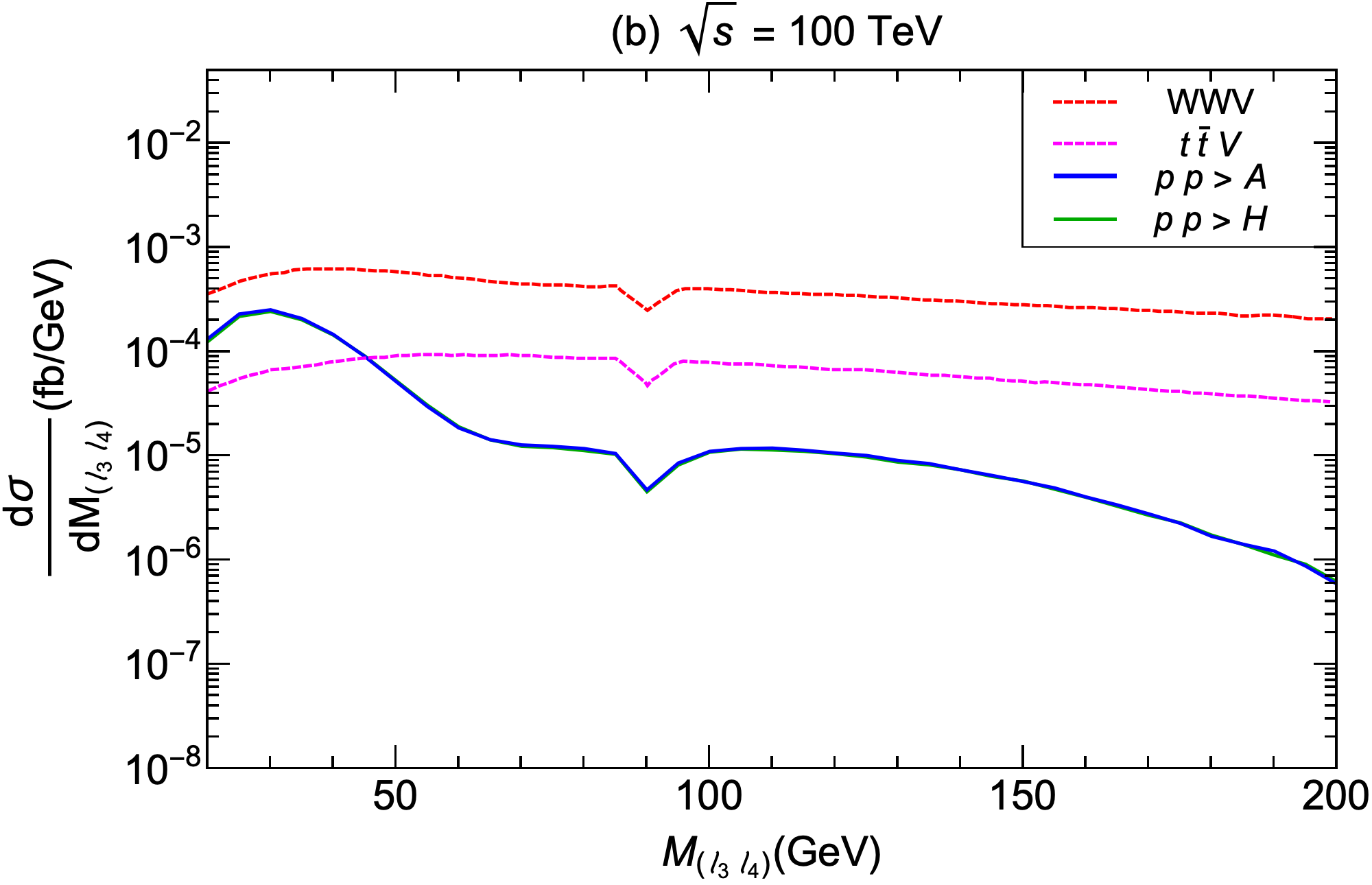}}\\ 
  {\includegraphics[width=0.48\textwidth]{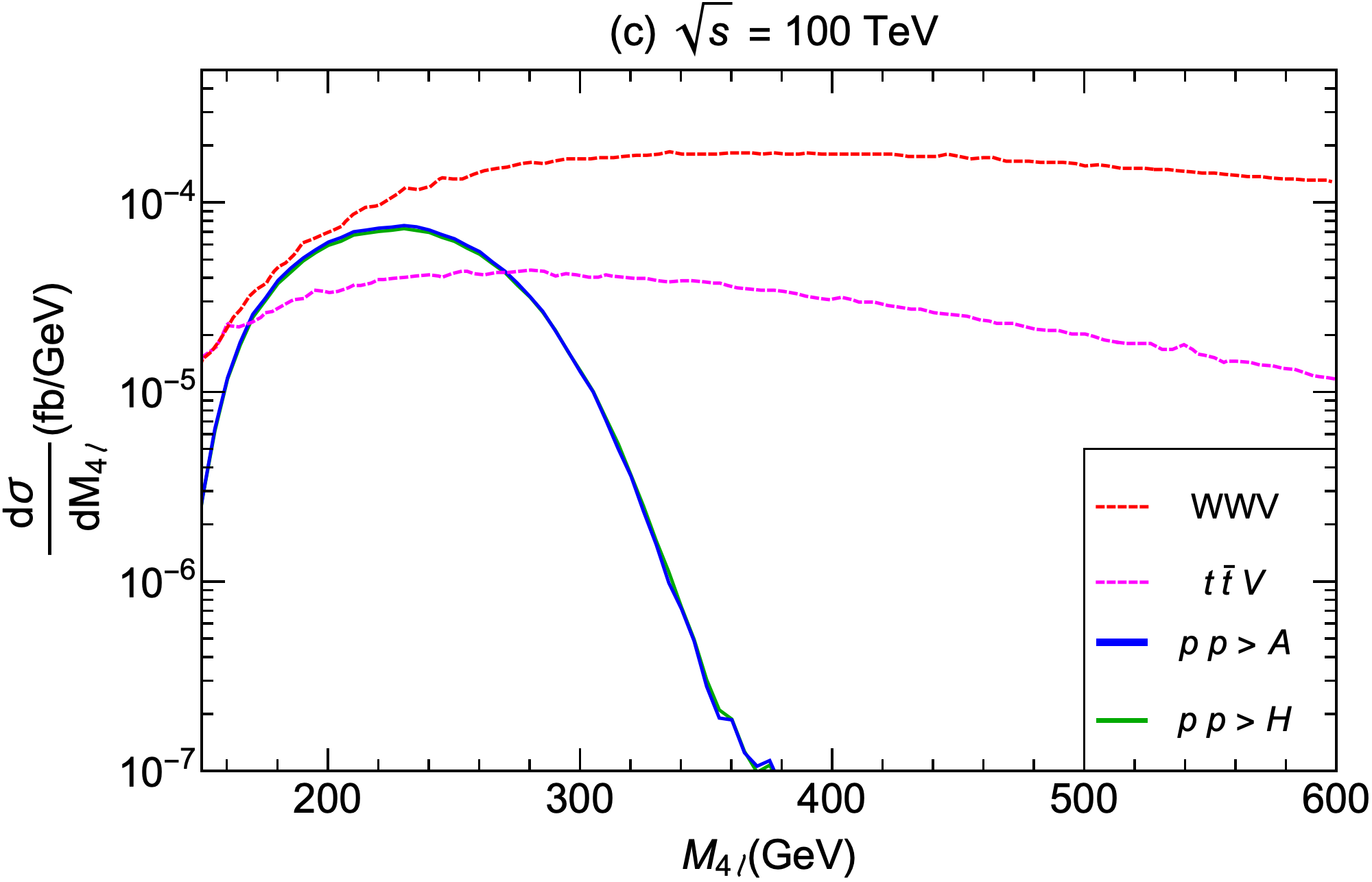}} 
   \hspace{1mm}
  {\includegraphics[width=0.48\textwidth]{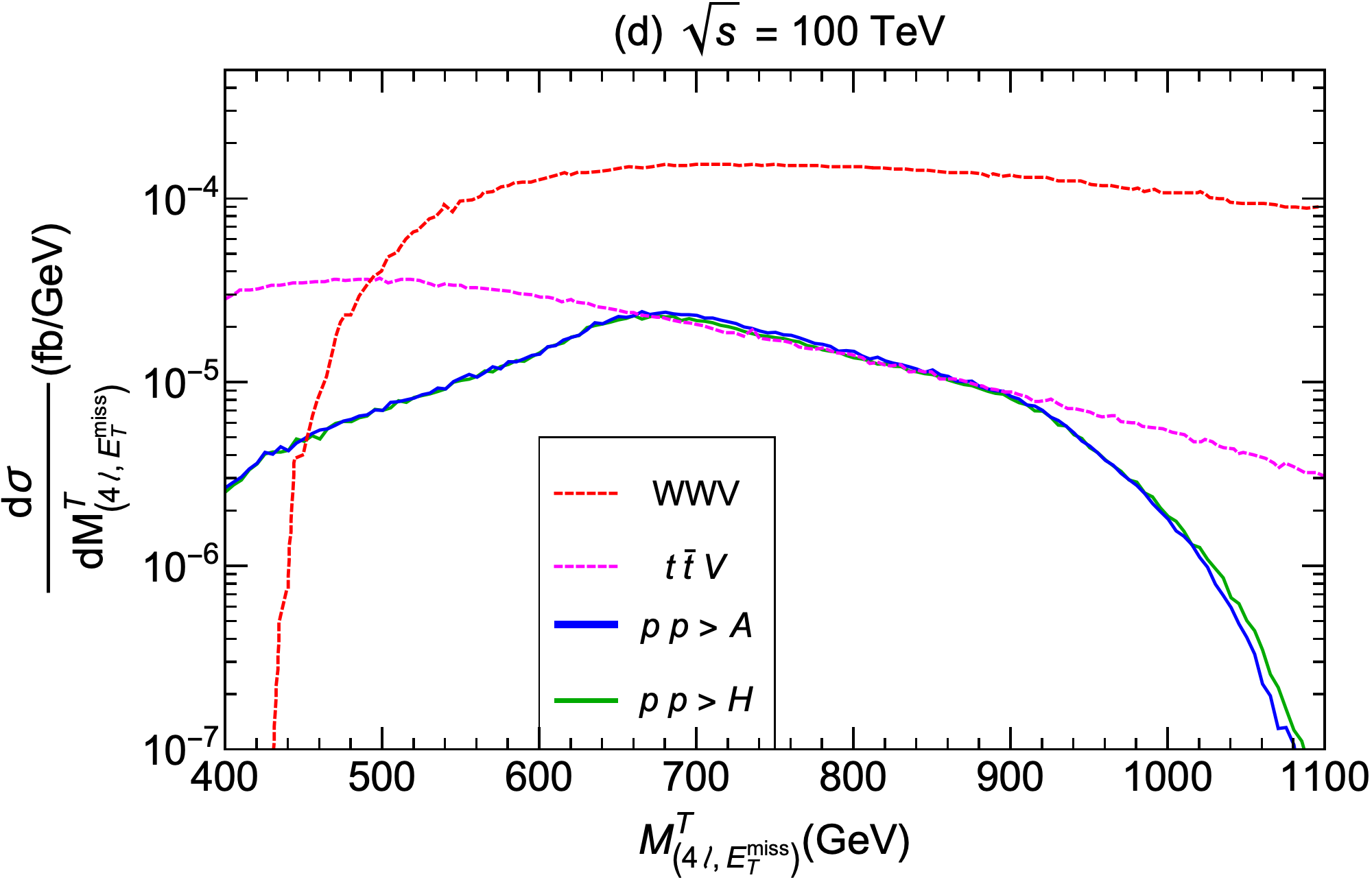}}\\
  \caption{Plots of the (a) invariant mass distribution 
    $M(\ell_1\ell_2)$ of the two leptons that form an invariant mass
    closest to $m_Z$,(b) invariant mass distribution of the remaining two
    leptons, $M(\ell_3,\ell_4)$, (c) invariant mass of the $4\ell$
    system, and (d) cluster transverse mass distribution of the
    $4\ell+\slashed{E}_T$ system, for the Higgs signal $(pp \to H,\ A
    \to 4\ell +\slashed{E}_T +X)$, after the cut set A defined in the
    text.  The corresponding contributions from the dominant physics
    backgrounds are also shown.  }
\label{fig:4}
\end{figure}

\begin{figure}[h!]
\centering
\includegraphics[width=0.68\textwidth]{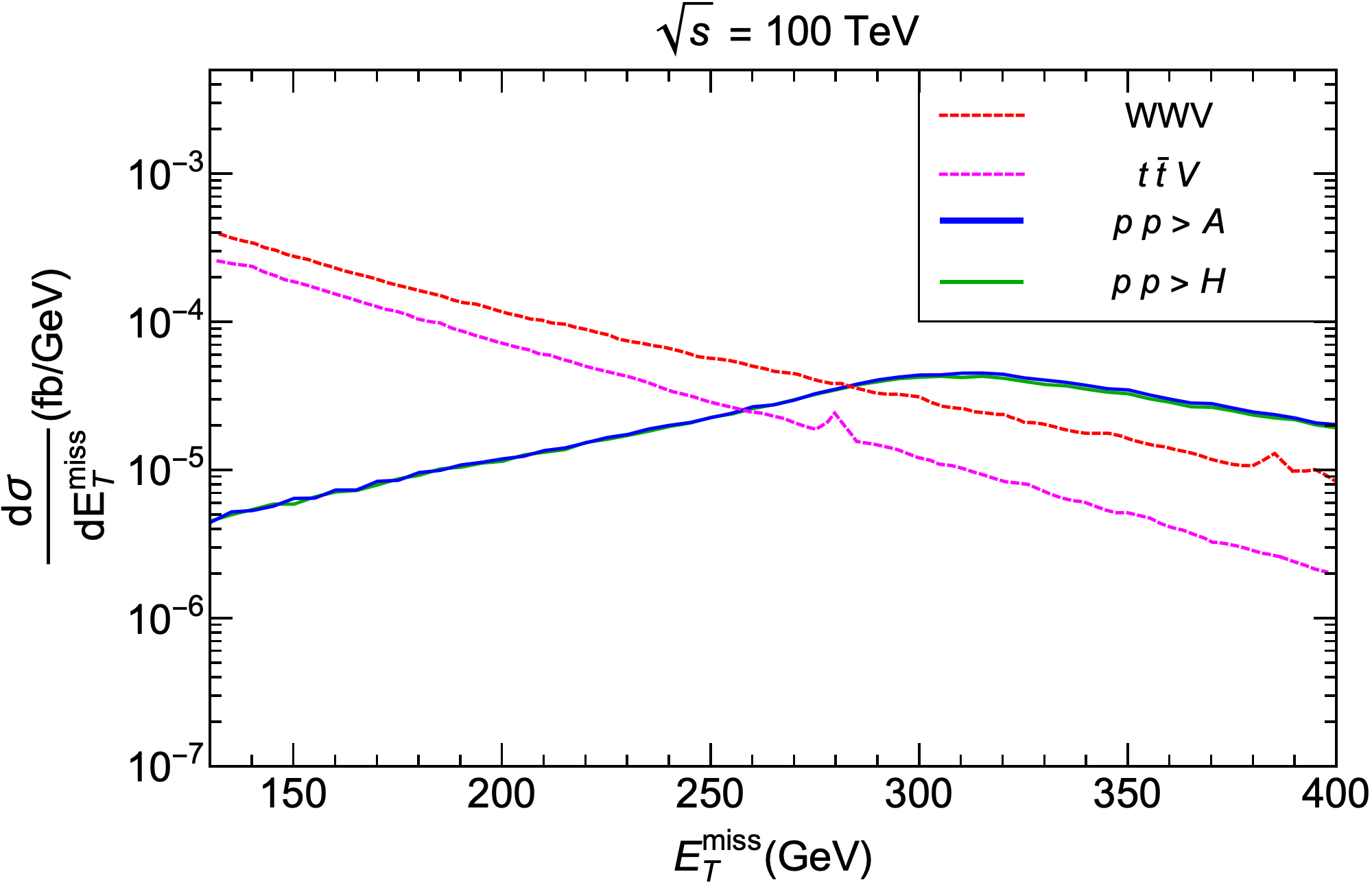}
\caption{The missing transverse energy $\slashed{E}_T$ distribution for the Higgs
  signal $(pp \to H,\ A \to 4\ell +\slashed{E}_T +X)$ after the cuts set
  A.  The corresponding contributions from the dominant physics
  backgrounds are also shown.}
\label{fig:5}
\end{figure}

After applying cut A, the mass distributions and $\slashed{E}_T$
distribution obtained (upon summing $b\bar{b}$ and $gg$ initiated
processes) are shown in Fig~\ref{fig:4} and \ref{fig:5},
respectively.

Since neutralinos and neutrinos escape detection (and so serve as
sources of missing energy) it is not possible to reconstruct the
invariant mass of $H$ or $A$ as a bump in the invariant mass of the
final state. We can, however, sharpen the signal by additional cuts.
Motivated by \cite{Aaboud:2018xdl}, we apply $\slashed{E}_{T} \geq 275$
GeV cut, since we have two neutralinos of mass $\sim$ 100 GeV in the
final state.  As can be seen from Figs~\ref{fig:4} and \ref{fig:5}, the
following mass cuts and $\slashed{E}_T$ cuts can reduce the SM
background very efficiently. Further cuts applied are :
\begin{itemize}
    \item We define $\ell_1$ and $\ell_2$ as the two leptons whose
      invariant mass is closest to $m_Z$ and require $|M(\ell_1,\ell_2)
      - m_Z| < 10$ GeV since the signal includes one $Z$
      boson;\footnote{Although we do not explicitly require it, for the
        most part, $\ell_1$ and $\ell_2$ have opposite sign and same
        flavour.}
    \item $10 < M(\ell_3,\ell_4) < 75$ GeV, where $\ell_3$ and $\ell_4$
      denotes the remaining leptons.
    \item $0.14 \ m_A < M(4\ell) < 0.34 \ m_A$
    \item $\slashed{E}_T > 275$ GeV.
\end{itemize}
Of course, since $m_A$ is not {\em a priori} known, the cut on
$M(4\ell)$ needs further explanation. Unless $m_A$ has already been
measured from studies of $A$ or $H$ decays via SM channels,
operationally, $m_A$ here refers to the upper end point of the signal
$M_T(4\ell,\slashed{E}_T)$ distribution shown in frame (d) of
Fig.~\ref{fig:4}, assuming that it can be experimentally
extracted.\footnote{We appreciate that the extraction of this end point
  may be very difficult. Since this is a first exploration of the
  $4\ell+\slashed{E}_T$ signal from the decay of heavy Higgs bosons in
  natural SUSY models, we do not attempt to explore the details of the
  end point determination, but simply assume that it can be extracted
  from the data.}  We note that the optimal choice of the $M(4\ell)$ cut
would only be weakly sensitive to the lightest neutralino mass for
$m_{A,H}\gg m_{\tz_1}$.  The cut set A, augmented by the cuts listed
above, is labeled as cut {\bf set B}.

\begin{figure}[h!]
\begin{center}
\includegraphics[width=0.48\textwidth]{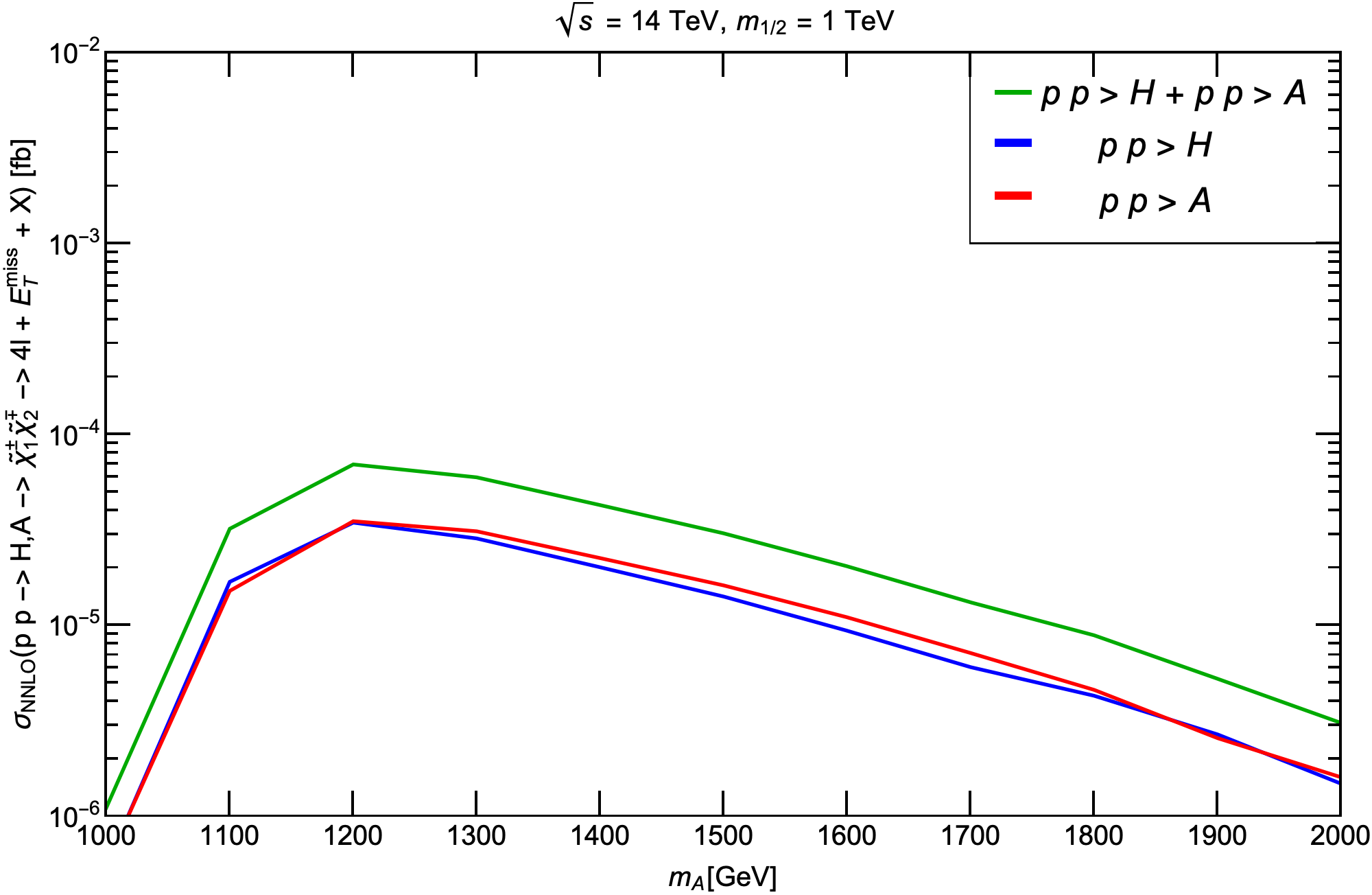}
\hspace{1mm}
\includegraphics[width=0.48\textwidth]{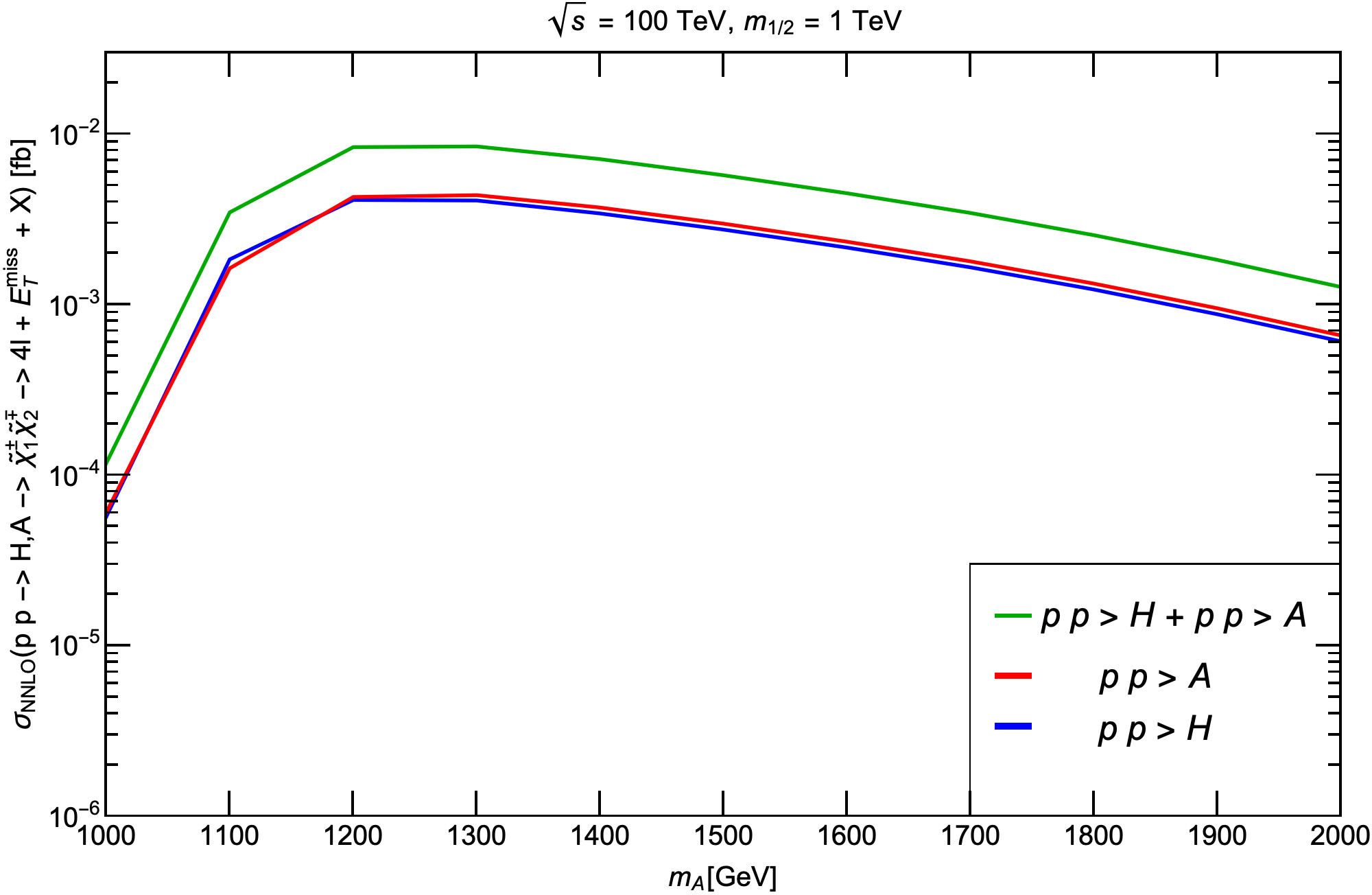}
\caption{NNLO Cross sections, $\sigma(A)$, $\sigma(H)$, and
  $\sigma(A)+\sigma(H)$ times the cascade decay branching fractions into
  the $4\ell +\eslt$ final state in fb vs. $m_A$ for (a) 14 TeV and (b)
  100 TeV, after the cut set B defined in the text.}
\label{fig:sig4lMETcuts}
\end{center}
\end{figure}

In Fig.~\ref{fig:sig4lMETcuts}, we show the signal cross section versus
$m_A$ after cuts B at (a) the HL-LHC, and (b) a 100~TeV $pp$
collider. We indeed see from frame (a) that for all values of $m_A$
the signal lies well  below the one event level. Although
perhaps only of academic interest, it is worth noting that a comparison
of this figure with Fig.~\ref{fig:sig4lMET}(a) shows that the signal
efficiency is $\sim$~5-10\% despite the requirement all four leptons are
required to have a
$p_T$ of at least 10~GeV. This is a reflection of the boost the
electroweakinos, and concomitantly the leptons, gain when they
originate in the decays of the heavy Higgs bosons. From
Fig.~\ref{fig:sig4lMETcuts}(b), we project that at the FCC or at the
SPPC with an integrated luminosity of 15~ab$^{-1}$, several tens of
signal events may be expected after  cuts B over most of the range of
$m_A$ in the figure.

\section{Discovery Potential with Cut-and-Count Analysis}
\label{sec:cuts}

In this section, we study
the discovery potential of the $4\ell +\eslt$ signal for
heavy Higgs bosons at a 100~TeV $pp$ collider using a traditional
cut-and-count analysis. 
To this end, we show in Table~\ref{tab:Cutbased}
our results for the signal after the cut set B 
for three benchmark points (BPs) with varying $m_A$ (with other parameters
fixed to their values in Table~\ref{tab:bm}), along with the main
sources of SM backgrounds. The subdominant background listed in the
fourth-last row is the combined background resulting from SM $Zh$ and from
$ZZV$ production.
\begin{table}[h!]
    \centering
    \begin{tabular}{|p{2cm}|p{3 cm}|p{3 cm}|p{ 3cm}|}\hline \hline 
    &  BP1 & BP2 & BP3 \\
   & $m_A$ \ = \ $1200\ \rm GeV$ & $m_A $ \ = \ $1400 \ \rm GeV$ & $m_A$ \ = \ $1600 \ \rm GeV$ \\ \hline
    $pp \to H$ & 4.12 $\times 10^{-3}$ & 3.45$\times 10^{-3}$ & 2.17 $\times 10^{-3}$ \\ \hline
    $pp \to A$ & 4.38 $\times 10^{-3}$ & 3.73 $\times 10^{-3}$ & 2.35 $\times 10^{-3}$ \\ \hline
    $W^+W^-\ell^+\ell^-$ & 7.13 $\times 10^{-3}$ & 7.23 $\times 10^{-3}$ & 6.18 $\times 10^{-3}$ \\ \hline
    $t\bar{t}\ell^+\ell^-$& 1.83 $\times 10^{-3}$ & 1.58 $\times 10^{-3}$ & 1.17 $\times 10^{-3}$ \\ \hline
    Z$\ell^+\ell^-\ell^+\ell^-$ & 1.38$\times 10^{-3}$ & 1.41 $\times 10^{-3}$ & 1.24$\times 10^{-3}$ \\ \hline
    $N_S$ & 127 & 108 &  68 \\ \hline
    $N_B$ & 155 & 153 & 129 \\ \hline
    $N_{ss}$ & 9.1 & 7.9 & 5.5 \\ \hline
    \end{tabular}
   
    \caption{The signal and SM backgrounds at a 100~TeV $pp$ collider
      for three benchmark points after the cut set B defined in the
      text.  All the cross sections are in fb. Here, $N_S$ is the total
      number signal events, combining both scalar and pseudo scalar and
      $N_B$ is the total number of background events and $N_{ss}$ is the
      statistical significance of the signal, all for an integrated
      luminosity of 15~$\rm ab^{-1}$. We have  all
      flavours of leptons ($e$ and $\mu$).}
\label{tab:Cutbased}
\end{table}

\begin{figure}[h!]
    \centering
    \includegraphics[width=0.68\textwidth]{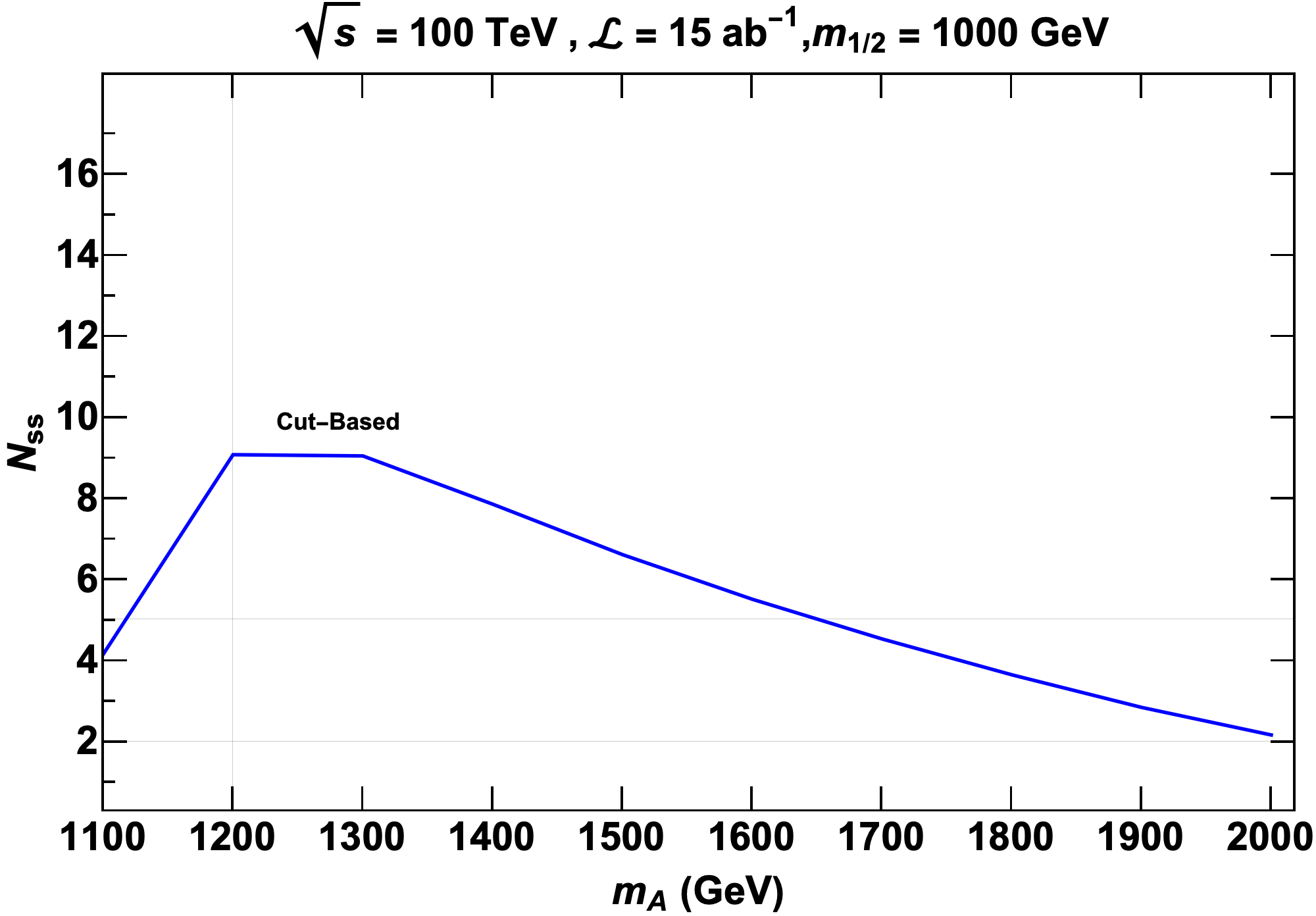}
    \caption{The signal significance $N_{ss}$ vs $m_A$ using a traditional
      cut-based analysis for $pp \to H+A \to 4\ell +\slashed{E}_T$ events
      at a 100~TeV $pp$ collider.}
    \label{fig:significance1}
\end{figure}

In Fig.~\ref{fig:significance1}, we present our estimates of
statistical significance \cite{stat},
$$N_{ss}\equiv \sqrt{(2\times(N_S + N_B)\ln(1 + N_S/N_B) - 2\times
  N_S)},$$ for 1100~GeV~$\leq m_A \leq 2000$~GeV. Our selection cuts
work well in removing a large part of the background.
We see that with a center of mass energy of 100 TeV and integrated luminosity of
$\mathcal{L}$ = 15 $\rm ab^{-1}$, we have enough events to claim a
$5\sigma$ discovery for $m_A\sim 1.1-1.65$ TeV.
We also obtain a 95\% CL exclusion limit for the
$H,\ A\to 4\ell +\eslt$ signal for values of $m_A$ extending out
as far as 2 TeV.

We now turn to an examination of whether we can use machine learning
techniques to suppress the background further and concomitantly increase
the reach.  In the next section, we study the  use of
boosted decision trees to further enhance the signal.

\section{Improvement with Boosted Decision Trees}
\label{sec:BDT}

We have just seen that the cut-based signal from heavy Higgs boson
decays via the $4\ell+\slashed{E}_T$ channel yields a
statistically significant discovery level over a limited 
range of $m_A$ values even at a 100~TeV $pp$ collider.
Of course, it is possible that this signal may be combined
with a signal from other channels to claim discovery over a wider range. 
The point of this study, however, is to examine how much improvement may
be possible without combining other channels if we go beyond the
traditional cut-based analysis which as we saw yields a
discovery significance of $N_{ss}> 5$ for $m_A\sim 1.1-1.65$ TeV
for $\sqrt{s}$ = 100 TeV and 15 ab$^{-1}$ of integrated luminosity.

It has been found that ML techniques can greatly improve the
signal-to-background discrimination and they are widely used by
experimental analyses.  In this section we use boosted decision trees (BDT)
for which algorithms are included in the ToolKit for MultiVariate
Analysis (TMVA)~\cite{TMVA}, a multivariate analysis package included
with ROOT.
For this study, we have used the following variables for training and testing,
\begin{itemize}
    \item The invariant mass $M(4\ell)$.
    \item The invariant masses $M(\ell_1, \ell_2)$ and $M(\ell_3,\ell_4)$ 
    \item $\slashed{E}_{T}$, missing transverse energy.
\end{itemize}

\begin{figure}[h!]
    \centering
    \includegraphics[width=0.48\textwidth]{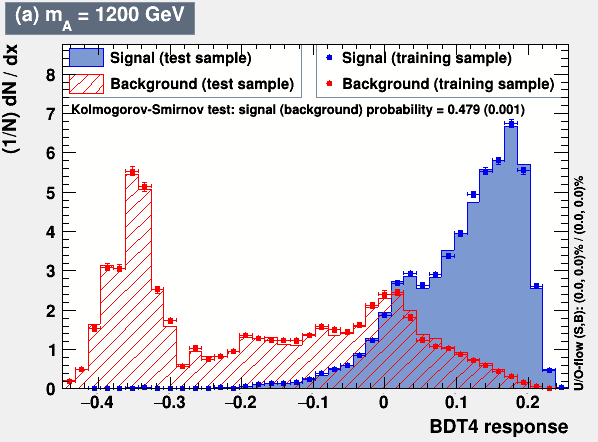}\\
    \includegraphics[width=0.48\textwidth]{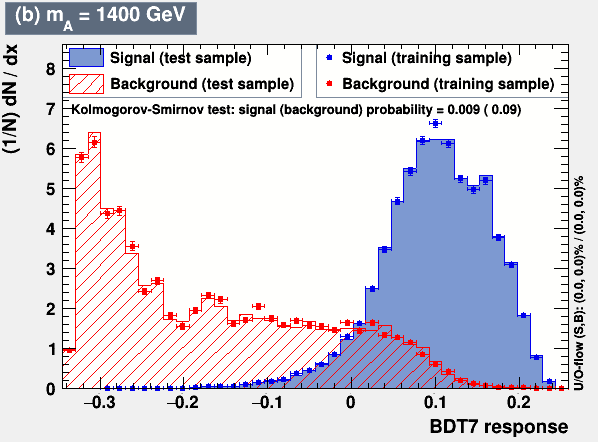}
    \hspace{1mm}
    \includegraphics[width=0.48\textwidth]{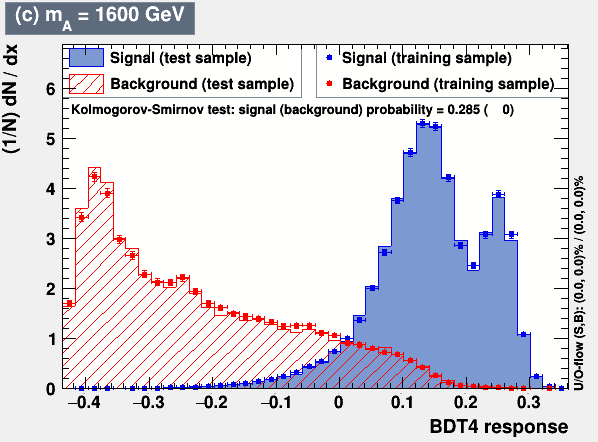}
    \caption{The BDT response for $m_A$ = (a) 1200, (b) 1400 and (c) 1600
      GeV. The BDT response of test points (solid) and training points
      (with error bar) is superposed in the figure.  }
    \label{fig:bdtresponse}
\end{figure}

We have generated signal files for each value of $m_A$ along with the
backgrounds at 100~TeV after applying the cut set {\bf B}, except that
we have now relaxed the cut on $\eslt$ to be $\eslt > 200$~GeV before passing
the samples for training and testing.
We train 400,000 signal events
and 400,000 background events for each channel.  We used the same number of
events for testing.  Figure~\ref{fig:bdtresponse} shows the BDT
response for three BPs with different $m_A$ values. 

In Table \ref{tab:bdt}, we present our estimate of $N_{ss}$ from the
BDT analysis for the same BP points as in Table~\ref{tab:Cutbased}.
We see that there is, indeed, a significant improvement over the previous
cut-based analysis.
\begin{table}[h!]
    \centering
    \begin{tabular}{c|c|c|c} \hline \hline
        Number of Events & pp $\to \phi^0$ & Total Background & $N_{ss}$ \\ \hline
          \multicolumn{4}{c}{BP1, $m_A = 1200\  \rm GeV$}\\\hline
          All mass cuts & 127 & 155 & 9.1 \\
          BDT cut  & 132 & 58 & 13.7 \\ \hline
            \multicolumn{4}{c}{BP2, $m_A = 1400\  \rm GeV$}\\\hline
          All mass cuts & 107 & 153 & 7.9 \\
          BDT cut  & 133 & 46 & 14.9 \\ \hline
            \multicolumn{4}{c}{BP3, $m_A = 1600\ \rm GeV$}\\\hline
          All mass cuts & 68 & 129 & 5.5 \\
          BDT cut  & 72 & 25 & 11.0  \\ \hline
         
    \end{tabular}
    \caption{A comparison between the cut based and BDT analyses for the
      three benchmark points introduced in the text.}
    \label{tab:bdt}
\end{table}

\begin{figure}
    \centering
    \includegraphics[width=0.48\textwidth]{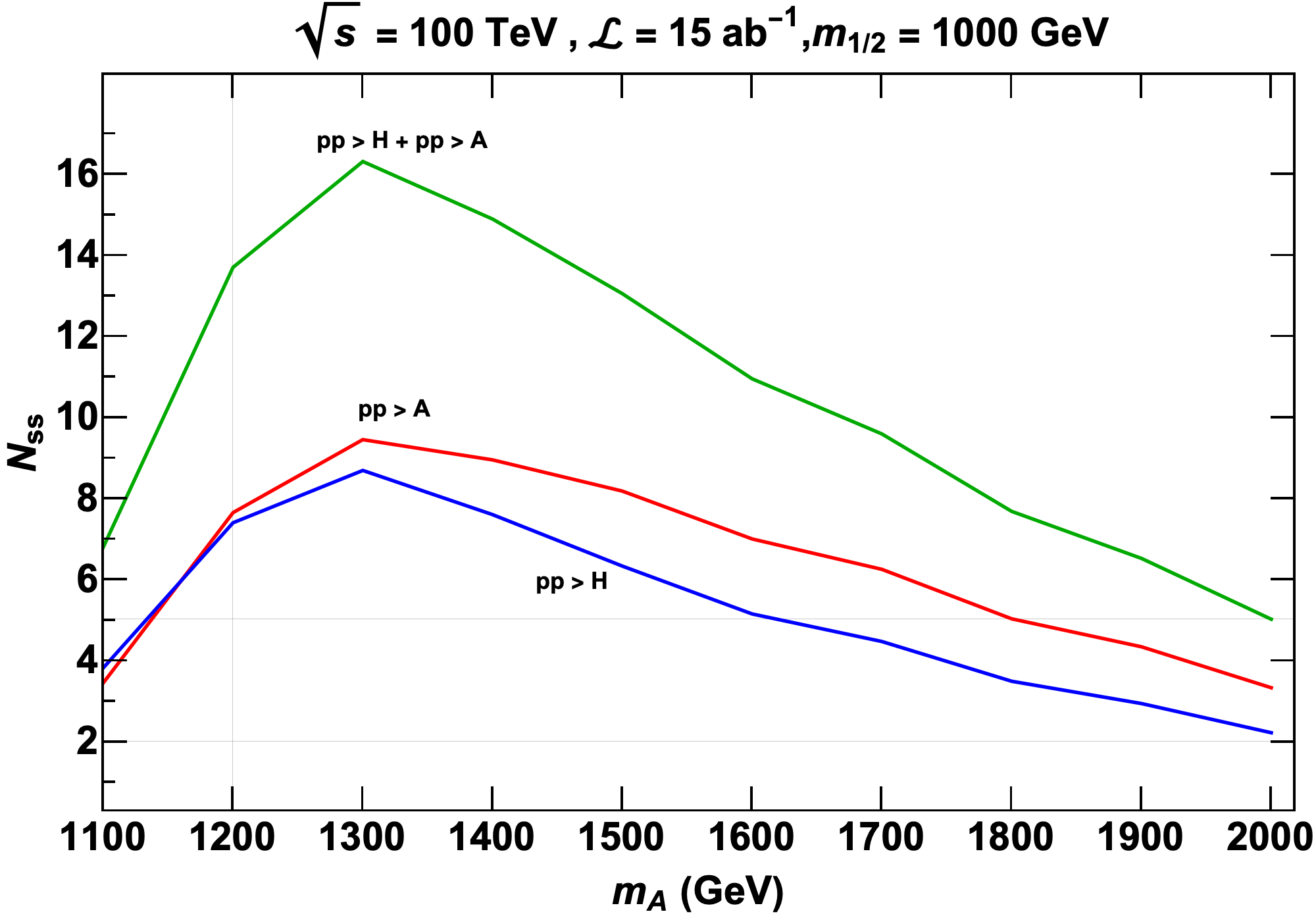}
    \caption{Statistical significance plots for the $H,\ A\to 4\ell +\eslt$ 
      signal at a 100~TeV hadron collider after the BDT analysis.
    }
    \label{fig:significance2}
\end{figure}

Fig.~\ref{fig:significance2} shows the individual contributions from
each of $H$ and $A$ for the BDT analysis along with the significance
from the combined $H$ and $A$ signal. This may be compared to the
significance shown in Fig.~\ref{fig:significance1} for the traditional
cut-and-count analysis.  We see that, by using the BDT analysis, we
would be able to discover $H$ and $A$ at the  $5\sigma$ level via
$H,\ A\to 4\ell +\eslt$ channel for $m_A\sim 1-2$ TeV -- a considerable
improvement in range of $m_A$ over the usual cut-based method!

\section{Conclusions}
\label{sec:conclude}

In this paper, we have examined heavy neutral Higgs boson discovery as
motivated by natural SUSY models with light higgsinos. In such models,
the heavy Higgs $H,\ A$ decays to electroweakinos are almost always open since the lightest higgsinos are expected to have masses below
$\sim 350$ GeV range whilst the $H$ and $A$ bosons can have TeV-scale
masses.  Since decays to pairs of higgsino-like states are dynamically
suppressed, our channel of primary interest is $H,\ A\to
\tchi_1^\pm\tchi_2^\mp$ decay, followed by $\tchi_2^\pm\to
Z\tchi_1^\pm$ followed by $Z\to \ell^+\ell^-$ and then each
$\tchi_1^\pm\to \ell^\pm\nu_{\ell}\tchi_1^0$.  Combining all flavours
of decays to $e$ and $\mu$ leads to a distinctive $H,\ A\to 4\ell
+\eslt$ signature for heavy Higgs boson decay to SUSY particles.  The
leptons from $\tchi_1^\pm$ decay are soft in the $\tchi_1^\pm$ rest
frame but are boosted to higher energies due to the large $m_{H,A}$
masses. Thus, we evaluated this signal channel against dominant SM
backgrounds for both HL-LHC and for FCC-hh or SPPC with $\sqrt{s}=100$
TeV, applying judicious cuts on various combinations of invariant
masses of the leptons, and also requiring $\slashed{E}_T> 275$~GeV.
Our selection requirements retain much of the signal while removing
the physics background efficiently.

In our analysis we have focused on production of the heavy Higgs bosons
with a mass ($m_H \simeq m_A$) between 1 TeV and 2 TeV.  While a
signal (in the $4\ell+\slashed{E}_T$ channel) is not likely to be
observable at HL-LHC, prospects are much better at FCC-hh or SPPC. The
best case for discovery is near $m_A \simeq 1.2-1.3$ TeV that has a balance
between kinematics of leptons in the final state and production cross
sections.  We note the following:
\begin{itemize}
\item A 100 TeV hadron collider offers promise to discover a heavy
  neutral Higgs boson via one of its dominant SUSY decay modes in
  natural SUSY models with a mass $\sim 1-2$ TeV.  With a conventional
  cut-based analysis, we are able to obtain a $N_{ss}>5$ statistical
  significance over a range $m_A\sim 1.1-1.65$ TeV.  We find though that
  a BDT analysis of the same signal can potentially improve the
  significance greatly giving $N_{ss}$ as high as 16 for
  $m_A\simeq~1.3$~TeV, and $N_{ss}>5$ over a range $m_A\sim 1-2$ TeV
  even via our proposed very difficult discovery channel.
\item For somewhat smaller values of heavy Higgs boson masses
  characterized by $m_A \alt 1$ TeV,
  the signal cross section is suppressed both by smaller
  branching ratio into the SUSY mode, and also by a smaller boost of the
  daughter EWinos which, in turn, reduces the efficiency with which the
  softer leptons pass the cuts. Nonetheless, the heavy neutral SUSY Higgs
  bosons should be detectable in this range via SM decay modes such as
  $H,\ A\to\tau\bar{\tau}$.
\item For increasing $m_A$ values beyond $\sim 1.3$ TeV,
  the Higgs production cross section
  becomes much smaller since the $gg$ and $b\bar{b}$ fusion production
  cross sections are increasingly suppressed.
\item We stress that we have focused only on the signal from a
  difficult SUSY decay mode of the heavy Higgs boson with an eye to
  assessing how ML techniques could serve to enhance difficult-to-see
  signals. Hence we have not examined the possibility of combining
  SUSY modes or whether the discovery of a heavy Higgs boson might be
  possible from a study of its SM decays.
\end{itemize}

For $m_A \simeq m_H$ significantly beyond 1 TeV and $\tan\beta\sim
10-50$, it may become increasingly challenging to search
for heavy Higgs bosons via their decays into SM particles due to the
diminished branching fractions to $b\bar{b}$ and $\tau\bar{\tau}$, once
the dominant SUSY decay channels become allowed.  The chargino and
neutralino discovery channel for heavy Higgs bosons at high energy
hadron colliders offers an important opportunity to discover the heavy
neutral Higgs bosons via their decay into EWinos.  An upgrade to a 100
TeV hadron collider seems essential for heavy Higgs $H$ and $A$
discovery via the natural SUSY $4\ell +\eslt$ channel.

\section*{Acknowledgement}

We thank an anonymous referee for useful suggestions.
This work was supported in part by the US Department of Energy,
 Office of High Energy Physics Grant No. DE-SC-0009956 and  DE-SC-0017647. The work of DS was supported in
part by the Ministry of Science and Technology (MOST) of Taiwan 
under Grant No. 110-2811-M-002-574, and work of RJ is supported by
 MOST 110-2639-M-002-002-ASP. 



\begin{thebibliography}{99}

\bibitem{higgs} G. Aad et al. [ATLAS Collaboration], Phys. Lett. B 716 (2012) 1; S. Chatrchyan et al. [CMS
Collaboration], Phys. Lett. B 716 (2012) 30,
%
\bibitem{ghp} E. Witten, Nucl. Phys. B 188, 513 (1981); R. K. Kaul, Phys. Lett. B 109, 19 (1982).
%
\bibitem{gcu} S. Dimopoulos, S. Raby and F. Wilczek, Phys. Rev. D 24 (1981) 1681; U. Amaldi, W. de Boer
and H. Furstenau, Phys. Lett. B 260, 447 (1991); J. R. Ellis, S. Kelley and D. V. Nanopoulos,
Phys. Lett. B 260 (1991) 131; P. Langacker and M. x. Luo, Phys. Rev. D 44 (1991) 817
%
\bibitem{top} L. E. Ibanez and G. G. Ross, Phys. Lett. B110, 215 (1982); K. Inoue et al. Prog. Theor. Phys. 68,
927 (1982) and 71, 413 (1984); L. Ibanez, Phys. Lett. B118, 73 (1982); H. P. Nilles, M. Srednicki
and D. Wyler, Phys. Lett. B 120 (1983) 346; J. Ellis, J. Hagelin, D. Nanopoulos and M. Tamvakis,
Phys. Lett. B125, 275 (1983); 
%
  L.~Alvarez-Gaume, J.~Polchinski and M.~B.~Wise,
  Nucl.\ Phys.\ B {\bf 221}, 495 (1983).
%
B. A. Ovrut and S. Raby, Phys. Lett. B 130 (1983) 277; for a review, see L. E. Ibanez
and G. G. Ross, Comptes Rendus Physique 8 (2007) 1013.
%
\bibitem{mhiggs} H. E. Haber and R. Hempfling, Phys. Rev. Lett. 66 (1991) 1815; J. R. Ellis, G. Ridolfi and
F. Zwirner, Phys. Lett. B 257 (1991) 83; Y. Okada, M. Yamaguchi and T. Yanagida, Prog.
Theor. Phys. 85 (1991) 1; For a review, see e.g. M. S. Carena and H. E. Haber, Prog. Part. Nucl.
Phys. 50 (2003) 63.
%
\bibitem{sven} S.~Heinemeyer, W.~Hollik and G.~Weiglein,
  Phys. Rept. \textbf{425} (2006), 265.
%
\bibitem{atlasg} The ATLAS collaboration [ATLAS Collaboration], ATLAS-CONF-2017-022.
%
\bibitem{cmsg} A. M. Sirunyan et al. [CMS Collaboration], Phys. Rev. D 97, no. 1, 012007 (2018); A. M. Sirunyan et al. [CMS Collaboration], Eur. Phys. J. C
77, no. 10, 710 (2017).
%
\bibitem{atlast} The ATLAS collaboration [ATLAS Collaboration], ATLAS-CONF-2017-037.
%
\bibitem{cmst} A. M. Sirunyan et al. [CMS Collaboration], arXiv:1706.04402 [hep-ex].
%
\bibitem{bg} R. Barbieri and G. F. Giudice, Nucl. Phys. B 306, 63 (1988).
%
\bibitem{dg} S.~Dimopoulos and G.~F.~Giudice,
  Phys. Lett. B \textbf{357} (1995), 573.
%
\bibitem{ac} G.~W.~Anderson and D.~J.~Castano,
  Phys. Rev. D \textbf{52} (1995) 1693.
%
\bibitem{unn1} R. Kitano and Y. Nomura, Phys. Rev. D 73 (2006) 095004;
M. Papucci, J. T. Ruderman and A. Weiler, JHEP 1209 (2012) 035.
%
\bibitem{unn} N. Craig, arXiv:1309.0528 [hep-ph].
%
\bibitem{dew} H.~Baer, V.~Barger and D.~Mickelson,
  Phys.\ Rev.\ D {\bf 88}, 095013 (2013);
A.~Mustafayev and X.~Tata,
  Indian J.\ Phys.\  {\bf 88} (2014) 991;
H.~Baer, V.~Barger, D.~Mickelson and M.~Padeffke-Kirkland,
  Phys.\ Rev.\ D {\bf 89}, 115019 (2014).
%
\bibitem{rns} H.~Baer, V.~Barger, P.~Huang, A.~Mustafayev and X.~Tata,
Phys. Rev. Lett. \textbf{109} (2012) 161802; 
H.~Baer, V.~Barger, P.~Huang, D.~Mickelson, A.~Mustafayev and X.~Tata,
Phys. Rev. D \textbf{87} (2013) no.11, 115028.
%
\bibitem{baerbook} H.~Baer and X.~Tata, {\it Weak Scale Supersymmetry: From Superfields to Scattering Events},
(Cambridge University Press, 2006).
%
\bibitem{upper} H.~Baer, V.~Barger and M.~Savoy,
Phys. Rev. D \textbf{93} (2016) no.3, 035016.
%
\bibitem{bounds} H.~Baer, V.~Barger, J. S. Gainer, P. Huang, M. Savoy,H. Serce and X. Tata, Phys.Lett. B774 (2017) 451.
%
\bibitem{bbbms} K.~J.~Bae, H.~Baer, V.~Barger, D.~Mickelson and M.~Savoy,
Phys. Rev. D \textbf{90} (2014) no.7, 075010.
%
\bibitem{bbdkt} H.~Baer, M.~Bisset, D.~Dicus, C.~Kao and X.~Tata,
Phys. Rev. D \textbf{47} (1993) 1062.
%
\bibitem{bbkt} H.~Baer, M.~Bisset, C.~Kao and X.~Tata,
  Phys. Rev. D \textbf{50} (1994) 316.
%

\bibitem{Bordry:2018gri}
F.~Bordry, M.~Benedikt, O.~Br\"uning, J.~Jowett, L.~Rossi, D.~Schulte, S.~Stapnes and F.~Zimmermann,
[arXiv:1810.13022 [physics.acc-ph]].

\bibitem{CEPC-SPPCStudyGroup:2015csa}
M.~Ahmad, D.~Alves, H.~An, Q.~An, A.~Arhrib, N.~Arkani-Hamed, I.~Ahmed, Y.~Bai, R.~B.~Ferroli and Y.~Ban, \textit{et al.}
IHEP-CEPC-DR-2015-01.

\bibitem{euro} [European Strategy~Group],
``2020 Update of the European Strategy for Particle Physics,''
doi:10.17181/ESU2020
%
\bibitem{isajet} F.~E.~Paige, S.~D.~Protopopescu, H.~Baer and X.~Tata,
  hep-ph/0312045.
%
\bibitem{nuhm2} D.~Matalliotakis and H.~P.~Nilles,
  Nucl.\ Phys.\ B {\bf 435} (1995) 115;
M.~Olechowski and S.~Pokorski,
  Phys.\ Lett.\ B {\bf 344} (1995) 201;
P.~Nath and R.~L.~Arnowitt,
  Phys.\ Rev.\ D {\bf 56} (1997) 2820;
J. Ellis, K. Olive and Y. Santoso, Phys. Lett. {\bf B539} (2002) 107;
J. Ellis, T. Falk, K. Olive and Y. Santoso, 
Nucl. Phys. {\bf B652} (2003) 259;
H.~Baer, A.~Mustafayev, S.~Profumo, A.~Belyaev and X. Tata, 
JHEP{\bf 0507} (2005) 065.
%
\bibitem{ATLAS-H} G.~Aad \textit{et al.} [ATLAS],
Phys. Rev. Lett. \textbf{125} (2020) no.5, 051801.
%
\bibitem{lhcsoftdilep} G.~Aad {\it et al.} (ATLAS Collaboration)
  Phys. Rev. D {\bf 101} (2020) 052005; A.~Tumasyan {\it et al.} (CMS
  Collaboration) arXiv:2111.06296 (2021).
%
\bibitem{bddt} H.~Baer, D.~Dicus, M.~Drees and X.~Tata,
Phys. Rev. D \textbf{36} (1987) 1363.
%
\bibitem{ssc1} J.~F.~Gunion, H.~E.~Haber, M.~Drees, D.~Karatas, X.~Tata, R.~Godbole and N.~Tracas,
Int. J. Mod. Phys. A \textbf{2} (1987) 1035.
%
\bibitem{gh3} J.~F.~Gunion and H.~E.~Haber,
Nucl. Phys. B \textbf{307} (1988) 445.
[erratum: Nucl. Phys. B \textbf{402} (1993), 569].
%
\bibitem{ghaber} K.~Griest and H.~E.~Haber,
Phys. Rev. D \textbf{37} (1988) 719.
%
\bibitem{kz} Z.~Kunszt and F.~Zwirner,
Nucl. Phys. B \textbf{385} (1992) 3.
%
\bibitem{djkz} A.~Djouadi, P.~Janot, J.~Kalinowski and P.~M.~Zerwas,
Phys. Lett. B \textbf{376} (1996) 220.
%
\bibitem{bbgh} V.~D.~Barger, M.~S.~Berger, J.~F.~Gunion and T.~Han,
Phys. Rept. \textbf{286} (1997) 1.
%
\bibitem{belanger} G.~Belanger, F.~Boudjema, F.~Donato, R.~Godbole and S.~Rosier-Lees,
Nucl. Phys. B \textbf{581} (2000) 3.
%
\bibitem{cdls} S.~Y.~Choi, M.~Drees, J.~S.~Lee and J.~Song,
Eur. Phys. J. C \textbf{25} (2002) 307.
%
\bibitem{css} C.~Charlot, R.~Salerno and Y.~Sirois,
  J. Phys. G \textbf{34} (2007) N1.
%
\bibitem{bisset} M.~Bisset, J.~Li, N.~Kersting, R.~Lu, F.~Moortgat and S.~Moretti,
JHEP \textbf{08} (2009) 037.
%
\bibitem{bbns} K.~J.~Bae, H.~Baer, N.~Nagata and H.~Serce,
Phys. Rev. D \textbf{92} (2015) no.3, 035006.
%
\bibitem{bbcc} R.~K.~Barman, B.~Bhattacherjee, A.~Chakraborty and A.~Choudhury,
Phys. Rev. D \textbf{94} (2016) no.7, 075013.
%
\bibitem{bagnaschi} E.~Bagnaschi, H.~Bahl, E.~Fuchs, T.~Hahn, S.~Heinemeyer, S.~Liebler, S.~Patel, P.~Slavich, T.~Stefaniak and C.~E.~M.~Wagner, \textit{et al.}
Eur. Phys. J. C \textbf{79} (2019) no.7, 617.
%
\bibitem{gls} S.~Gori, Z.~Liu and B.~Shakya,
JHEP \textbf{04} (2019) 049.
%
\bibitem{abgkk}A.~Adhikary, B.~Bhattacherjee, R.~M.~Godbole, N.~Khan and S.~Kulkarni,
JHEP \textbf{04} (2021) 284.
%
\bibitem{higgspro} S.~Dittmaier \textit{et al.} [LHC Higgs Cross Section Working Group],
doi:10.5170/CERN-2011-002
[arXiv:1101.0593 [hep-ph]].
%

\bibitem{Harlander:2012pb}
R.~V.~Harlander, S.~Liebler and H.~Mantler,
Comput. Phys. Commun. \textbf{184} (2013) 1605.
\bibitem{Harlander:2016hcx}
R.~V.~Harlander, S.~Liebler and H.~Mantler,
Comput. Phys. Commun. \textbf{212}, 239-257 (2017).
\bibitem{Harlander:2002wh}
R.~V.~Harlander and W.~B.~Kilgore,
Phys. Rev. Lett. \textbf{88}, 201801 (2002).
\bibitem{Harlander:2003ai}
R.~V.~Harlander and W.~B.~Kilgore,
Phys. Rev. D \textbf{68}, 013001 (2003).
\bibitem{Aglietti:2004nj}
U.~Aglietti, R.~Bonciani, G.~Degrassi and A.~Vicini,
Phys. Lett. B \textbf{595}, 432-441 (2004).
\bibitem{Bonciani:2010ms}
R.~Bonciani, G.~Degrassi and A.~Vicini,
Comput. Phys. Commun. \textbf{182}, 1253-1264 (2011).
\bibitem{Degrassi:2010eu}
G.~Degrassi and P.~Slavich,
JHEP \textbf{11}, 044 (2010).
\bibitem{Degrassi:2011vq}
G.~Degrassi, S.~Di Vita and P.~Slavich,
JHEP \textbf{08}, 128 (2011).
\bibitem{Degrassi:2012vt}
G.~Degrassi, S.~Di Vita and P.~Slavich,
Eur. Phys. J. C \textbf{72}, 2032 (2012).
\bibitem{Harlander:2005rq}
R.~Harlander and P.~Kant,
JHEP \textbf{12}, 015 (2005).
\bibitem{Chetyrkin:2000yt}
K.~G.~Chetyrkin, J.~H.~Kuhn and M.~Steinhauser,
Comput. Phys. Commun. \textbf{133}, 43-65 (2000).

%
%
\bibitem{mcfm}
R.~Boughezal, J.~M.~Campbell, R.~K.~Ellis, C.~Focke, W.~Giele, X.~Liu, F.~Petriello and C.~Williams,
Eur. Phys. J. C \textbf{77}, no.1, 7 (2017).
%
  \bibitem{ATLAS:2013-004} The ATLAS collaboration[ATLAS Collaboration],
    ATLAS-PHYS-PUB-2013-004.
%
\bibitem{Aaboud:2018xdl} 
  M.~Aaboud {\it et al.} [ATLAS Collaboration],
  JHEP {\bf 1810} (2018) 180.
%
\bibitem{stat} G. Cowan,K.~Cranmer,E.~Gross and O.~Vitells,
  Eur. Phys. Jour. C 
  {\bf 71} (2011) 1554.
%
\bibitem{TMVA} A.~Hocker, P.~Speckmayer, J.~Stelzer, J.~Therhaag, E.~von Toerne, H.~Voss, M.~Backes, T.~Carli, O.~Cohen and A.~Christov, \textit{et al.}
[arXiv:physics/0703039 [physics.data-an]].
%

\end{thebibliography}
\end{document}